%% file: tildeLmuLtau_paper.tex
\documentclass[a4paper,11pt]{article}
\pdfoutput=1 % if your are submitting a pdflatex (i.e. if you have
\usepackage{jheppub} % for details on the use of the package, please
\input{PreRamble}

\newcommand{\beq}{\begin{equation} }
\newcommand{\eeq}{\end{equation}}
\newcommand{\bi}{\begin{itemize} }
\newcommand{\ei}{\end{itemize} }

\newcommand{\rep}[1]{\mathbf{#1}}
\newcommand{\repbar}[1]{\overline{\mathbf{#1}}}

\newcommand{\andeq}{\quad \mathrm{and} \quad}

\renewcommand{\L}{\mathcal{L}}
\newcommand{\LL}{\mathrm{L}}
\newcommand{\RR}{\mathrm{R}}
\newcommand{\U}{\mathrm{U}}
\newcommand{\SU}{\mathrm{SU}}
\newcommand{\eminus}{\vcenter{\hbox{\scalebox{0.6}[1]{$ - $}}}}	%Narrow minus signed (for e.g. negative exponents)

\newcommand{\sscript}[1]{{\scriptscriptstyle \mathrm{#1}}}
\newcommand{\RK}{$ R_{K^{(\ast)}} $\xspace}
\newcommand{\Umt}{$ \U(1)_{L_\mu\eminus L_\tau} $\xspace}
\newcommand{\gmu}{$ (g-2)_\mu $\xspace}

\definecolor{Red}{rgb}{1.,0.,0.}
\definecolor{Grn}{rgb}{0.,0.75,0.}
\definecolor{Blu}{rgb}{0.,0.,1.}

\definecolor{red}{rgb}{0.6,.0706,.1373}
\definecolor{blue}{rgb}{0,0.396,0.741}
\definecolor{orange}{rgb}{0.9,0.4,0}

\allowdisplaybreaks

\begin{document}

% more complex case: 4 authors, 3 institutions, 2 footnotes
\author[1]{Admir Greljo,}
\author[1]{Peter Stangl,}
\author[1]{Anders Eller Thomsen,}
\author[2]{and Jure Zupan}

\affiliation[1]{Albert Einstein Center for Fundamental Physics, Institute for Theoretical Physics, University of Bern, CH-3012 Bern, Switzerland}
\affiliation[2]{Department of Physics, University of Cincinnati, Cincinnati, Ohio 45221,USA}

% e-mail addresses: one for each author, in the same order as the authors
\emailAdd{admir.greljo@unibe.ch}
\emailAdd{stangl@itp.unibe.ch}
\emailAdd{thomsen@itp.unibe.ch}
\emailAdd{zupanje@ucmail.uc.edu}

\title{\mathversion{boldsans}
On $(g-2)_\mu$ From Gauged $\U(1)_X$
}

\abstract{
We investigate an economical explanation for the \gmu anomaly with a neutral vector boson from a spontaneously broken $\U(1)_X$ gauge symmetry. The Standard Model fermion content is minimally extended by 3 right-handed neutrinos.
Using a battery of complementary constraints, we perform a thorough investigation of the renormalizable, quark flavor--universal, vector-like $\U(1)_X$ models, allowing for arbitrary kinetic mixing. Out of 419 models with integer charges not greater than ten, only 7 models are viable solutions, describing a narrow region in model space. These are either $L_\mu-L_\tau$  or models with a ratio of electron to baryon number close to $-2$. The key complementary constraints are from the searches for nonstandard neutrino interactions. Furthermore, we comment on the severe challenges to chiral $\U(1)_X$ solutions and show the severe constraints on a particularly promising such candidate.
}

\maketitle

\newpage

%%%%%%%%%%%%%%%%%%%%%%%%%%%%%%%%%%%
\section{Introduction}
\label{sec:intro}
%%%%%%%%%%%%%%%%%%%%%%%%%%%%%%%%%%%

Measurements~\cite{Bennett:2006fi,Muong-2:2021ojo} of the muon anomalous magnetic moment, $(g-2)_\mu$, are showing a combined $4.2\sigma$ deviation from the consensus Standard Model~(SM) prediction~\cite{Aoyama:2020ynm,Colangelo:2020lcg,aoyama:2012wk,Aoyama:2019ryr,czarnecki:2002nt,gnendiger:2013pva,davier:2017zfy,keshavarzi:2018mgv,colangelo:2018mtw,hoferichter:2019gzf,davier:2019can,keshavarzi:2019abf,kurz:2014wya,melnikov:2003xd,masjuan:2017tvw,Colangelo:2017fiz,hoferichter:2018kwz,gerardin:2019vio,bijnens:2019ghy,colangelo:2019uex,Blum:2019ugy,colangelo:2014qya}.\footnote{The issue of the SM prediction may not be completely settled. The lattice QCD determination of the hadronic vacuum polarization contribution to $(g-2)_\mu$ by the BMW group  reduces the tension between the experimental world average and the SM $(g-2)_\mu$ prediction to less than two standard deviations~\cite{Borsanyi:2020mff}. It is expected that other lattice groups will be able to weigh in on this issue in the near- to medium-term future. }
This may be pointing to the existence of new physics~(NP) coupling to muons. Such a possibility is especially intriguing in light of similar discrepancies in the $b \to s \mu^+ \mu^-$ observables:
angular distributions~\cite{LHCb:2020lmf,LHCb:2020gog}; branching ratios~\cite{LHCb:2020zud,LHCb:2021awg,LHCb:2021vsc,LHCb:2014cxe,LHCb:2015wdu,LHCb:2016ykl,LHCb:2021zwz}; and the theoretically very clean lepton flavor universality~(LFU) ratios, \RK~\cite{LHCb:2017avl,LHCb:2021trn} (for recent global fits see, e.g.,~\cite{Altmannshofer:2021qrr,Geng:2021nhg,Alguero:2021anc,Hurth:2021nsi,Ciuchini:2020gvn}). Global significance of the NP hypothesis in $b \to s \ell^+ \ell^-$ decays, including the look-elsewhere effect, was estimated to be $4.3\sigma$ in Ref.~\cite{Isidori:2021vtc}.

Any NP model explaining the muon anomalies faces nontrivial experimental constraints. Especially stringent are the constraints from lepton flavor--violating~(LFV) transitions, such as $\mu\to e\gamma$.
While the $(g-2)_\mu$ anomaly requires a relatively low \textit{effective} NP scale $\Lambda_{22} \simeq \SI{15}{TeV}$,
the bound on the flavor-changing $\mu \to e \gamma$ transition requires  $\Lambda_{12 (21)}  \gtrsim \SI{3600}{TeV}$~\cite{TheMEG:2016wtm}.\footnote{Both transitions are due to dipole moment operators, $\mathcal{L}_{\rm eff} \supset - { e \,v}\, \bar \ell^i_{\LL} \sigma^{\mu \nu} \ell^j_{\RR } F_{\mu \nu}/{(4 \pi \Lambda_{ij})^2} + {\rm H.c.}$, where $v = \SI{246}{GeV} $ is the electroweak vacuum expectation value. The NP contribution to $(g-2)_\mu$ is due to a flavor-conserving dipole, $i=j=2$, whereas $\mu\to e\gamma$ decay is from a flavor-changing dipole, $i=1, j=2$ or $i=2, j=1$.}
Viable NP explanations of $(g-2)_\mu$, therefore, must have highly suppressed LFV effects (as discussed in, e.g., Refs.~\cite{Isidori:2021gqe,Calibbi:2021qto}).

Strict flavor alignment between the dipole operator and the charged lepton mass matrix may well point to the existence of a new underlying symmetry, which we assume to be a spontaneously broken $ \U(1)_X$ gauge group. The neutral gauge vector boson associated with the $ \U(1)_X$ is then a candidate for an explanation of the \gmu.
One well-studied example of this scenarios is \Umt~\cite{He:1990pn,He:1991qd,Baek:2001kca,Ma:2001md,Harigaya:2013twa,Altmannshofer:2014pba,Altmannshofer:2019zhy,Crivellin:2016ejn,Crivellin:2015mga,Crivellin:2018qmi,Altmannshofer:2014cfa,Altmannshofer:2015mqa,Asai:2018ocx}. The \Umt gauge symmetry  forces the dimension-4 charged lepton Yukawa interactions to be diagonal, ensuring an accidental $\U(1)^3$ lepton flavor symmetry. The $L_\mu - L_\tau$ gauge boson $X_\mu$ with mass $m_{X}\in [10,\, 210]$\,MeV and a coupling to muons in the range $g_X\in [0.4,\, 1]\times 10^{-3}$ gives
a one-loop contribution to \gmu of the right size to account for the experimental anomaly while not conflicting with any of other measurements~\cite{Altmannshofer:2014pba,Bauer:2018onh,Escudero:2019gzq,Amaral:2021rzw,Greljo:2021npi}.

There are two immediate questions pertaining to the anomaly-free $\U(1)_X$ gauge extensions of the SM:
\begin{enumerate}[i)]
\item Is \Umt the only phenomenologically viable possibility that can explain the $(g-2)_\mu$ anomaly?
\item If there are alternative models, can these be experimentally disentangled from \Umt?
\end{enumerate}

In this paper, we systematically explore the two questions, building on our previous work, Ref.~\cite{Greljo:2021npi}.
The main result is that the only significant deviation of $ L_{\mu} - L_{\tau} $ allowed by data is gauge groups where $ L_e \simeq -2 B $ and the kinetic mixing with the photon approximately cancels the electron charge.
This conclusion is rather nontrivial since there exists an extensive list of precise experimental probes constraining complementary combinations of  $\U(1)_X$ gauge boson couplings, resulting in a limited number of phenomenologically viable possibilities.

The main working assumptions for our analysis are \emph{i)} the SM is minimally extended by three right-handed neutrinos and a flavor--non-universal $\U(1)_X$ gauge symmetry and \emph{ii)} the theory is anomaly-free.
The NP models also contain an SM singlet scalar field $\phi$ charged under $\U(1)_X$, whose vacuum expectation value (VEV) $v_\phi$ breaks the $\U(1)_X$.
The solutions to the \gmu anomaly typically require $ v_\phi $ around the electroweak scale if muons and $\phi$ carry ${\mathcal O}(1)$  $\U(1)_X$ gauge charges.
The details about the source of $\U(1)_X$ symmetry breaking are not important for most of the phenomenology discussed in this paper, so our results do not change if a different SM-singlet condensate breaks $\U(1)_X$ (or several condensates). Actually, viable neutrino masses and mixings typically require more than one scalar field.

Requiring the ratios between (non-zero) $\U(1)_X$ charges for the chiral fermions to be at most ten gives roughly $21.5$ million inequivalent integer charge assignments, up to flavor permutation~\cite{Allanach:2018vjg}. We restrict our analysis to a subset of these, the 255 \textit{quark flavor--universal} vector-like $\U(1)_X$ charge assignments~\cite{Greljo:2021npi}. Taking into account the flavor permutations among charged SM fermions this gives 419 phenomenologically distinct $\U(1)_X$ models. Here, we assumed that the sterile neutrinos are heavy enough so as not to affect the low-energy phenomenology, and, thus, different charge assignments for sterile neutrinos lead to the phenomenologically equivalent $\U(1)_X$ model in this counting. In Section~\ref{sec:model-II}, we furthermore comment on the 21 chiral $\U(1)_X$ charge assignments.
The choice of quark flavor universality of $\U(1)_X$ charges is phenomenologically well motivated, since flavor--non-universal $\U(1)_X$ charges of quarks lead to dangerous flavor-changing neutral currents (FCNCs).
As illustrated in Ref.~\cite{Greljo:2021npi} by a benchmark example, even for the second-safest option, a third-quark-family model with down-alignment, the CKM rotation induced up-sector FCNCs effectively rule out the parameter space preferred by \gmu.\footnote{Similarly, the anomalous $\U(1)_X$  extensions of the SM with a light $X_\mu$ gauge boson are severely constrained due to the $1/m_X^2$ enhanced rates for rare FCNC decays. The $XWW$ anomalous triple gauge couplings induce axial $X \bar d_i d_j$ flavor-violating  couplings, which give rise to enhanced decays involving the longitudinal mode of the $X$ boson (see for example~\cite{Dror:2017nsg}).
}

While quark flavor--universal models avoid FCNCs, they are still severely constrained through a combination of other measurements:
\begin{enumerate}[i)]

\item neutrino trident constraints~\cite{Altmannshofer:2014pba,Mishra:1991bv,NuTeV:1999wlw,Altmannshofer:2019zhy};

\item electroweak precision observables~\cite{ParticleDataGroup:2020ssz,Haller:2018nnx,ALEPH:2005ab};

\item neutrino oscillation constraints on nonstandard neutrino interactions~(NSI)~\cite{Esteban:2018ppq,Coloma:2020gfv};

\item measurements of coherent neutrino scattering on nuclei~\cite{Freedman:1973yd,Drukier:1984vhf,COHERENT:2017ipa};

\item the Borexino measurement of the cross section for the elastic scattering of $^{7}$Be solar neutrinos on electrons~\cite{Bellini:2011rx,Borexino:2017rsf} and other elastic neutrino-electron scattering experiments~\cite{TEXONO:2009knm,Beda:2009kx,CHARM-II:1993phx,CHARM-II:1994dzw};

\item searches for new light resonances~\cite{Ilten:2018crw,Bauer:2018onh}.

\end{enumerate}
In Ref.~\cite{Greljo:2021npi}, we studied the implications of these measurements for several selected benchmarks. In this manuscript, we go well beyond the initial analysis of Ref.~\cite{Greljo:2021npi} and assess the constraints for the complete set of $419$ distinct vector-like $ \U(1)_X $ models, focusing on the parameter space region relevant for explaining $(g-2)_\mu$. In several instances our phenomenological analysis applies to the complete class of quark flavor--universal vector-like charge assignment, even allowing for arbitrarily large charge assignments. We also comment on the phenomenology of chiral charge assignments, working out the details of the $\tilde L_{\mu-\tau}$ model in which the couplings of $X_\mu$ to electrons are purely axial, thus, eliminating the very strict constraints on NSI from neutrino oscillations.

In the bulk of the paper, the analysis is kept as general as possible. In particular,  we do not impose any requirements regarding possible connections with the $b\to s\ell\ell$ anomalies. This does not mean that such connections do not exist. On the contrary, lepton flavor--non-universal $\U(1)_X$ gauge symmetries
can further support
solutions of the $ B $-physics anomalies that rely on tree-level exchanges of leptoquarks (LQ). In general, TeV-scale LQs tend to excessively violate the accidental symmetries of the SM---baryon and individual lepton number symmetries---all of which are exquisitely tested experimentally. Charging the LQ under a flavor--non-universal $\U(1)_X$ gauge symmetry can reinstate the accidental symmetries while keeping the contributions to $B$-anomalies intact. Prominent examples of such mediators are the muoquarks~\cite{Greljo:2021xmg,Davighi:2020qqa,Hambye:2017qix,Davighi:2022qgb,Heeck:2022znj}, LQs charged under a $\U(1)_X$ such that they carry global muon and baryon numbers. Both of these remain accidentally conserved  at dimension 4, as they are in the SM. Out of 255+21 quark flavor universal charge assignments, 252+21 satisfy the muoquark criteria for the scalar weak triplet mediator~\cite{Greljo:2021npi}.

The paper is organized as follows. In Section~\ref{sec:framework}, we introduce the anomaly-free $\U(1)_X$ models and the parameter space relevant for \gmu bounded by cosmology ($m_X \gtrsim 10$\,MeV) and perturbative unitarity ($m_X \lesssim 1$\,TeV). Section~\ref{sec:constraints} contains the discussion of the experimental constraints relevant for vector-like quark universal $\U(1)_X$ models. The phenomenological implications of these constraints are presented in Section~\ref{sec:pheno:implications}: In Section~\ref{sec:generic}, we use the neutrino trident production, combined with the $Z$-pole constraints, to set a robust upper limit on the $X$ boson mass ($m_X \lesssim 4$\,GeV) for all renormalizable models introduced in Section~\ref{sec:framework}. In Section~\ref{sec:global:constr}, we perform a global analysis of experimental constraints and identify seven viable vector-like models that can explain the $(g-2)_\mu$ anomaly.
Section \ref{sec:model-II} contains a brief discussion of chiral models and a detailed phenomenological analysis of the most promising example, the $\tilde L_{\mu-\tau}$ model. Possible connections with the $B$ anomalies are discussed in Section~\ref{sec:B-anomalies}, while Section~\ref{sec:conc} contains conclusions. The details on the calculation of neutrino oscillation bounds are given in Appendix~\ref{app:NSI}, while Appendix~\ref{app:global:chi2} contains details on the construction of the global $\chi^2$ function used in the analysis.

%%%%%%%%%%%%%%%%%%%%%%%%%%%%%%%%%%%

\section{Model framework}
\label{sec:framework}

We start by reviewing the salient features of the gauged $ \U(1)_X $ models and how these can address the \gmu.

\subsection{Field content and $\U(1)_X$ charges}

\label{sec:models}
The models we consider have the SM matter content extended by three sterile neutrinos $N_i$ while the SM gauge group is enlarged by an additional $\U(1)_{X}$ factor.
After electroweak symmetry breaking (EWSB), the part of the Lagrangian relevant to the phenomenology at low energies ($m_X \ll m_Z$)
is given by\footnote{In Section~\ref{sec:EWP}, we write up the underlying theory in the unbroken-phase of the SM to allow for all values of $m_X$ and be able to describe electroweak precision tests.}
\begin{equation} \label{eq:L:benchmark}
    \mathcal{L} \supset - \tfrac{1}{4} F_{\mu\nu}^2 - \tfrac{1}{4} X_{\mu\nu}^2 + \tfrac{1}{2} \varepsilon \, F_{\mu\nu} X^{\mu\nu} +  \tfrac{1}{2} m_X^2 X_\mu^2 + e A_\mu J_\sscript{EM}^\mu + g'_X X_\mu J_X^\mu,
\end{equation}
where $F_{\mu\nu}$ and $X_{\mu\nu}$ are the electromagnetic and $\U(1)_X$ field strength tensors, $\varepsilon$ is the kinetic mixing parameter, $ J^\mu_{\sscript{EM}} $ the electromagnetic current, and $
J_X^\mu= \sum_f x_f \overline{f} \gamma^\mu f$  the current associated with $ \U(1)_X $, where $x_f$ are the $ \U(1)_X $ charges for the chiral field $f$.
The kinetic mixing term $\frac{\varepsilon}{2} \, F_{\mu\nu} X^{\mu\nu} $
can be removed by performing a non-unitary transformation of the Abelian gauge fields~\cite{Holdom:1985ag}, after which the $\U(1)_{X}$  Lagrangian is given by setting  $\varepsilon \to 0$,   $ g'_X \to g_X = g'_X/\sqrt{1- \varepsilon^2} $ in Eq.~\eqref{eq:L:benchmark} and shifting the charges according to
    \begin{equation} \label{eq:low-energy_charges}
    x_f \to x_f + \frac{\varepsilon}{\sqrt{1-\varepsilon^2}} \dfrac{e}{g_X} Q_f^\sscript{EM}
    \end{equation}
in the expression for $J_X^\mu$
($ Q_f^\sscript{EM} $ is the electric charge of the SM fermion $f$).

\subsection{Charge assignments}
\label{sec:charges}

The $ \U(1)_X$ charges of quarks are assumed to be universal, such that the quark Yukawa interactions are given by the usual dimension-4 operators. Without loss of generality, we can set the $ \U(1)_X$ charge of the SM Higgs to zero, $ x_H = 0$. This leaves 276 inequivalent charge assignment, not counting flavor permutations, with integer charges for the SM fields in the range from -10 to 10, as listed in~\cite{Greljo:2021xmg}. Shifting all the $x_f$ charges by a multiple of the hypercharge $Y_f$ gives physically equivalent models that would have $ x_H \neq 0$, see Appendix~A.1 of Ref.~\cite{Greljo:2021xmg}.

The above \textit{quark flavor--universal} $ \U(1)_X$ charge assignments fall into one of two categories. The 255 charge assignment in the \textit{vector category} have vector-like $\U(1)_X$ charges for both the quarks and the charged leptons, $x_{L_i} = x_{E_i}$, $i=1,2,3$. The charged lepton masses are thus also generated via dimension-4 SM Yukawa interactions. Viable neutrino masses and mixings usually require additional $\U(1)_X$-breaking scalars (SM singlets), which lead to Majorana masses through $N_i N_j \phi_{ij}$ interactions. We assume that the sterile neutrinos are heavy enough that they are not relevant for the low energy phenomenology. Taking into account the flavor permutations of the vector-like $ \U(1)_X$ charge assignments  this leaves us with 419 phenomenologically distinct vector-like $\U(1)_X$ models.\footnote{Permutations of the right-handed charge assignments (in the charge lepton mass basis) are in principle possible. However, this requires higher-dimensional effective operators to generate the SM charged lepton Yukawa couplings while at the same time the renormalizable operators need to be suppressed. We ignore this rather artificial possibility. }
The possible charge assignments for the models in the vector category are given by~\cite{Altmannshofer:2019xda,Greljo:2021xmg}
\begin{equation} \label{eq:Tx}
    x_f
    =   c_{e} {L_{e}}+c_{\mu} {L_{\mu}}+c_{\tau}
    {L_{\tau}} -\left(\tfrac{c_{e}+c_{\mu}+c_{\tau}}{3}\right)
    {B} +\sum_i c_{N_i} L_{N_i},
\end{equation}
where $\{B,{L_{e}},{L_{\mu}},{L_{\tau}}\}$ are the usual values of baryon and lepton numbers for the SM fermion $f$, while $L_{N_i}$ are the right-handed neutrino numbers ($N_i$ are not charged under $B,L_{\ell_i}$). The coefficients $\{c_{e},c_{\mu},c_{\tau}, c_{N_1},c_{N_2},c_{N_3}  \}$ in Eq.~\eqref{eq:Tx} need to satisfy the Diophantine equations~\cite{Dobrescu:2020evn,Allanach:2018vjg}
($ c_e+c_\mu+c_\tau=\sum_i c_{N_i}$ and $c_e^3+c_\mu^3+c_\tau^3=\sum_i c_{N_i}^3$) giving a 4 parameter family of models valid beyond the restriction to charges less than 10.
In particular, there are valid solutions for any set of $c_e, c_\mu, c_\tau $: $ c_{N_i} = (-c_e,\, -c_\mu,\, -c_\tau) $.
The parameterization in Eq.~\eqref{eq:Tx} will be exploited in Section~\ref{sec:NSI} to
asses NSI constraints on a large set of models.

The charge assignments $ c_{N_1} = - c_{N_2} $, $ c_{N_3}= c_{e,\mu,\tau} =0 $ (and permutations thereof) correspond to the dark photon type solutions to \gmu~\cite{Pospelov:2008zw}.
In this case, the $ X $ couplings to the SM fermions are exclusively due to kinetic mixing with the photon---typically radiatively induced from $ X $ interacting with some hidden sector particles also charged under the SM gauge group. This scenario has been ruled out as the solution of the  \gmu anomaly, both when $ X $ decays visibly~\cite{NA482:2015wmo} or invisibly~\cite{NA64:2016oww,BaBar:2017tiz}, with the possible exception of a combined decay~\cite{Mohlabeng:2019vrz}.

There are additional 21 charge assignments for which
there is no permutation such that $x_{L_i} = x_{E_i}$ for every $i=1,2,3$. These models constitute the \textit{chiral category} and are listed in Section~2.2.2 of~\cite{Greljo:2021xmg}. We examine their phenomenology and possible relevance for $(g-2)_\mu$ separately, in Section~\ref{sec:model-II}. In chiral models at least some of the charged lepton Yukawa couplings to the Higgs are forbidden at dimension four, leading to Yukawa matrices of rank less than three. Hence, some of the charged lepton masses are generated through higher-dimension operators.
We assume that these operators are either generated by integrating out vector-like fermions
(which do not change the conditions for anomaly cancellation),
or by integrating out heavy scalars.
We discuss this in more detail for the $\tilde L_{\mu-\tau}$ model, including the phenomenological constraints, in Section~\ref{sec:model-II}.

\subsection{Mater unification}
\label{sec:unification}

Interestingly, some of the above $\U(1)_X$ extensions of the SM can be unified into a larger semi-simple gauge group at higher energies. As an example consider the vector category charge assignments with $x_{\ell_i} = x_{N_i}$. Ref.~\cite{Davighi:2022qgb} showed that in that case  the $\U(1)_Y \times \U(1)_X$ can be unified into $\SU(2)_R \times \U(1)_{B-L} \times \SU(3)_{{\rm lep}}$. Starting from this semi-simple group the unification can proceed further (see Figure 1 of Ref.~\cite{Davighi:2022qgb}).

\subsection{Explanation of \gmu}
\label{sec:g-2}
A massive vector $X_\mu$ coupling to muons with interaction
\beq
\label{eq:L:amu}
\L \supset
g_X\, \overline{\mu}\, \slashed X (q_{V} - q_{A} \gamma_5) \mu,
\eeq
 gives rise to a one-loop radiative correction to the anomalous magnetic moment of the muon:
    \beq
   \begin{split}
    \label{eq:amu_NP}
    \Delta a_\mu = \frac{g_X^2 }{8\pi^2}
    \begin{dcases}
    q_V^2 -2\, r_\mu^2\, q_A^2,  & m_X \ll m_\mu, \\
    \frac{2}{3} r_\mu^2 \left[  q_V^2 - 5\, q_A^2 \right], &  m_X \gg m_\mu,
    \end{dcases} \,
    \end{split}
    \eeq
where $ r_\mu = m_\mu / m_X $. For the full 1-loop expressions see, e.g., Refs.~\cite{Jackiw:1972jz,Jegerlehner:2009ry,Greljo:2021npi}. Correction~\eqref{eq:amu_NP} is of the right sign to explain the difference between the measured and predicted anomalous magnetic moment of the muon,
$ \Delta a_\mu
=   a^{\rm avg}_\mu - a^\sscript{ SM}_\mu
=   \left( 251 \pm 59 \right) \times 10^{-11}$~\cite{Muong-2:2021ojo}, if $X_\mu$ couples mainly vectorially to muons.

Requiring that $\Delta a_\mu$ is explained by a massive vector $X_\mu$ at one-loop order translates to an upper bound on the $\U(1)_X$ breaking VEV
$\sqrt{2} \langle \phi \rangle=m_X/(g_X x_\phi)<  260\text{~GeV} \times (q_V/x_\phi) $, which is saturated for  $q_A=0$.
Perturbative unitarity, $g_X q_V \leq \sqrt{4\pi}$, then implies an upper bound on the $X_\mu$ mass, $m_X \lesssim 1.0$ TeV
(see Ref.~\cite{Capdevilla:2021kcf} for a more detailed  discussion). This is a rather restrictive bound, the origin of which can be traced back to the fact that in the one-loop $X_\mu$ contribution to $(g-2)_\mu$, the required chirality flip
necessarily occurs on the muon leg and, thus, is suppressed by the small muon mass.

%%%%%%%%%%%%%%%%%%%%%
\subsection{Cosmology}
\label{sec:cosmo}

A robust lower limit on the gauge boson mass $m_X$ follows from the agreement between observations of primordial light element abundances and the predictions within standard cosmology Big Bang Nucleosynthesis (BBN). The gauge coupling $g_X$ required to explain the \gmu anomaly is large enough that the gauge boson $X_\mu$ efficiently thermalises with the SM plasma in the early universe, i.e., for temperatures $T > m_X$. A light $X_\mu$ contributes with additional relativistic degrees of freedom during BBN, changing the predictions for light element abundances. This is avoided for $m_X \gtrsim 10$\,MeV, in which case $X_\mu$ decays quickly, before the onset of BBN. The precise bound on $m_X$ was derived for the $\U(1)_{L_\mu - L_\tau}$ model in Ref.~\cite{Kamada:2018zxi} (see also~\cite{Escudero:2019gzq}), and holds approximately also for all the other $\U(1)_X$ models considered here.

\section{Experimental constraints}
\label{sec:constraints}
As a next step in the analysis, we discuss all the various constraints on the $ \U(1)_X $ models. They include EW precision tests, a variety of neutrino interactions, and finally resonance searches.

\subsection{Electroweak precision tests}\label{sec:EWP}
After EWSB the $\U(1)_X$ gauge boson $ X_\mu$ mixes both with the photon, $A_\mu$, and the $Z$ boson. The mixing with photon is important for low energy constraints, while the mixing with the $Z$ is severely constrained by the electroweak precision tests.

\subsubsection*{Electroweak symmetry breaking}
In the models we consider, the source of the EWSB is the same as in the SM---the  VEV of the SM Higgs.
Above the electroweak scale the kinetic mixing Lagrangian between $X_\mu$ and the photon, Eq.~\eqref{eq:L:benchmark}, is replaced by the kinetic kinetic mixing between $ X_\mu $ and $ B_\mu $, parametrized by the parameter $ \varepsilon_Y $:
	\begin{equation} \label{eq:lag_EWX}
	\L \supset - \tfrac{1}{4} B_{\mu\nu}^2 - \tfrac{1}{4} X^2_{0,\mu\nu} + \tfrac{1}{2} \varepsilon_Y \, B_{\mu\nu} X_0^{\mu\nu} - \tfrac{1}{4} (W_{\mu\nu}^a)^2 + |D_\mu H|^2 +  \tfrac{1}{2} m_{X_0}^2 X_{0,\mu}^2,
	\end{equation}
where $D_\mu H=(\partial_\mu +i g_X' x_H X_{0,\mu})H$. In writing the above Lagrangian we remain agnostic about the origin of the  $ X $ boson mass. The mass term $ m_{X_0} $ could be due to a VEV of a single SM singlet scalar or due to a more complicated condensate in the SM singlet sector.

The $ \U(1)_X $ charge of the SM Higgs, $ x_H $,  can be absorbed in the other parameters  by performing the shift
    \begin{equation}
        \big\{g'_X,\, \varepsilon_Y \big\} \longrightarrow \frac{1}{\sqrt{1+2\varepsilon_Y\, \xi + \xi^2}} \big\{g'_X,\, \varepsilon_Y + \xi \big\},
    \end{equation}
where  $\xi = 2x_H {g'_X}/{g_1}$.
This shift changes the charges $x_f$ of the matter fields, $f$, by
    \begin{equation}
    x_f \longrightarrow x_f - 2x_H Y_f,
    \end{equation}
with $ Y_f $ the hypercharge of fermion $f$. Without loss of generality, we can, therefore, take $x_H=0$, which is what we will assume from now on.

Expressing $ B_\mu $ and $ W_\mu^3 $ in Eq.~\eqref{eq:lag_EWX} in terms of the photon, $A_\mu$, and the $Z$ boson field, $Z_\mu$, gives the kinetic mixing Lagrangian after EWSB:
	\begin{equation}
	\begin{split}
	\label{eq:L:mix:EWSB}
	\L \supset &- \tfrac{1}{4} \big(A_{\mu\nu}^2 + X^{2}_{0,\mu\nu}+ Z_{0,\mu\nu}^{2}\big) + \tfrac{1}{2} m_{Z_0}^2 \! Z_{0,\mu}^2+ \tfrac{1}{2} m_{X_0}^2 X_{0,\mu}^{2}\\
	&+ \tfrac{1}{2} \varepsilon_Y \! \big(c_w A_{\mu\nu} - s_w Z_{0,\mu\nu} \big) X_0^{\mu\nu},
	\end{split}
	\end{equation}
where $ m_{Z_0} = g_Z \langle H \rangle / \sqrt{2} $ is the SM  $ Z $ mass. In Eq.~\eqref{eq:L:mix:EWSB} we do not write down the terms involving dynamical Higgs field. We also used the shorthand notation $s_w=\sin\theta_w$, $c_w=\cos\theta_w$, and $t_w=\tan \theta_w$, where $\theta_w$ is the weak mixing angle.

A simultaneous diagonalization of the kinetic and mass terms is achieved by a combination of a non-unitary field redefinition and a rotation,
	\begin{equation} \label{eq:field_shift}
	\begin{pmatrix}
	A^\mu \\ Z_0^\mu \\ X_0^\mu
	\end{pmatrix}\! =
	\begin{pmatrix}
	1 & \phantom{1/ r_\varepsilon}& c_w \hat \varepsilon_Y \\
	 & 1 & \eminus s_w \hat \varepsilon_Y \\
	 & & \Delta_\varepsilon \\
	\end{pmatrix} \!
	\begin{pmatrix}
	1 & & \\
	& c_\theta &  s_\theta \\
	& \eminus s_\theta & c_\theta \\
	\end{pmatrix}\!
	\begin{pmatrix}
	A^\mu \\ Z^\mu \\ X^\mu
	\end{pmatrix} ,
	\end{equation}
where $ \Delta_\varepsilon^2 = 1/(1 - \varepsilon_Y^2) $.
The mixing angle $ \theta $ between $ Z $ and $ X $ is given by
\begin{equation}
\label{eq:t:theta}
    t_\theta = \begin{dcases}
   \frac{s_w \hat \varepsilon_Y }{1-r_X^2}, & |s_w \hat \varepsilon_Y| < |1- r_X^2|, \\
    \frac{1-r_X^2}{s_w \hat \varepsilon_Y}, & |s_w \hat \varepsilon_Y| > |1- r_X^2|,
    \end{dcases}
    \end{equation}
where
	\begin{equation}
	r_X = \dfrac{m_{X}}{m_{Z_0}}, \andeq
	\hat \varepsilon_Y = \dfrac{\varepsilon_Y}{\sqrt{1 - \varepsilon_Y^2}},
	\end{equation}
with $ m_X $ the physical $ X $ boson mass.

The diagonalization of the gauge fields changes their effective currents. From Eq.~\eqref{eq:field_shift}, we find that
	\beq \label{eq:currents}
	\L \supset  e J_A^\mu A_\mu + g_Z J_Z^\mu Z_{0,\mu} + g_X' J_X^\mu X_{0,\mu} =  e J_A^\mu A_\mu  + g_Z  J_{Z,{\rm eff}}^\mu Z_{\mu} + g_X J_{X,{\rm eff}}^\mu X_{\mu},
	\eeq
with
\begin{align}
\label{eq:tildeJZ}
	J_{Z,{\rm eff}}^\mu &= c_\theta \! \left[ J^\mu_Z - {c_w\, t_\theta\, \hat \varepsilon_Y} J_Y^\mu - t_\theta \dfrac{g_X}{g_Z} J_X^\mu \right]\!,
	\\
	\label{eq:tildeJX}
	J_{X,{\rm eff}}^\mu &= J_X^\mu +  c_w \,c_\theta \hat \varepsilon_Y \dfrac{ e}{g_X} J_A^\mu	+ c_\theta \dfrac{\, g_Z}{g_X}\! \left(t_\theta - s_w \hat  \varepsilon_Y \right)\! J_Z^\mu ,
\end{align}
where $ g_X = c_\theta \Delta_\varepsilon  g'_X  $ is the physical $ X $ gauge coupling.

\subsubsection*{$ Z$ mass constraint}
\label{sec:Tparam}
The kinetic mixing between $ X_\mu $ and $ B_\mu $ introduces a mass mixing between $ Z $ and $ X $ gauge bosons, resulting in a shift of the physical $ Z $ mass $ m_Z$ and corrections to $Z$ couplings to SM fermions. While a global fit to the electroweak precision data would be required to capture \emph{all} the resulting experimental constraints on $ X_\mu$--$B_\mu $ mixing, it suffices for our purposes to consider just the $T$ parameter (marginalized over other electroweak observables), mainly because the measurements of SM fermion couplings to $Z$ have larger relative errors, see also discussion in Ref.~\cite{Efrati:2015eaa}, as well as Refs.~\cite{Babu:1996vt,Babu:1997st,Cassel:2009pu,Curtin:2014cca}, and Section~\ref{sec:LFU:Z} below.

We find that
	\begin{equation}
	\rho_0 = \frac{m_{Z,\sscript{SM}}^2}{ m_{Z,\mathrm{obs}}^2} \simeq \frac{m_{Z_0}^2}{ m_{Z}^2} = \dfrac{1- r_X^2}{1- r_X^2 + s_w^2 \hat \varepsilon_Y^2},
	\end{equation}
where $m_{Z,\sscript{SM}}$ is the SM prediction for the $Z$ mass (from the $W$ mass including radiative corrections), whereas $m_{Z,\mathrm{obs}}$ is the measured $Z$ mass so that in the SM $\rho_0=1$. Since we are interested in the NP constraints, we can use the tree-level relations, giving the second (approximate) equality. $\rho_0$ is related to the oblique parameter $T$ through $\rho_0-1=1/(1-\hat \alpha(m_Z) T)-1\simeq \hat \alpha(m_Z) T$, where experimentally from the electroweak fit $ T= \num{0.03(12)} $, when the $S,U$ parameters are allowed to float freely~\cite{ParticleDataGroup:2020ssz} (see also~\cite{Baak:2014ora}).
The $ Z $ mass shift in our model results in
	\begin{equation} \label{eq:T_contribution}
	T= - \dfrac{1}{\alpha(m_Z)} \dfrac{s_w^2 \hat \varepsilon_Y^2}{1 - r_X^2 + s_w^2 \hat \varepsilon_Y^2}.
	\end{equation}
In the limit of small $ X $ mass and small kinetic mixing parameter, we have
    \begin{equation}
	T= - \dfrac{s_w^2}{\alpha}\varepsilon^2_Y \left[1 +\mathcal{O}\big(\varepsilon_Y^2,\, m_X^2/m_Z^2 \big)^2 \right] \simeq - 30\, \varepsilon^2_Y.
    \end{equation}
This provides a relatively weak bound for the region of light $X$ masses ($ m_X< 2 m_\mu $)
    \beq
    \label{eq:epsY:bound}
    |\varepsilon_Y| < 0.076,
    \eeq
well above the typical expectation $\varepsilon_Y \sim {\mathcal O }(e g_X /16\pi^2)$ for radiatively induced kinetic mixing.
Nevertheless, the $T$ parameter bound in Eq.~\eqref{eq:epsY:bound} is phenomenologically quite important, since it does not allow for arbitrarily large kinetic mixings. Combined with constraints from neutrino trident production, it translates to a model-independent requirement that $X$ needs to be lighter than a few GeV (see Section~\ref{sec:NTP} for details).

For $X$ masses comparable to $m_Z$, the  bound in Eq.~\eqref{eq:epsY:bound} becomes more stringent and then progressively weaker for $m_X\gg m_Z$. Throughout this region, the mixing angle $\theta$ between $X$ and $Z$ is given by the upper expression in Eq.~\eqref{eq:t:theta}, except for the very small region where $X$ and $Z$ are almost mass degenerate, such that $ |s_w \hat \varepsilon_Y| > |1- r_X^2|$. In this mass degenerate regime, the constraint on the $T$ parameter leads to
  \begin{equation}
   |s_w \hat \varepsilon_Y| < \frac{\alpha |T|}{1- \alpha |T| }  \implies |\varepsilon_Y| < \num{4.7e-3}.
    \end{equation}
As anticipated, in this region, the constraint on   $ |\varepsilon_Y|$ is significantly more stringent than it is for the light $X$ mass limit, Eq.~\eqref{eq:epsY:bound}.
From the $ T $ bound, it follows that this case is only relevant for $ |1- r_X^2| <\num{2.2e-3} $, i.e., when the $X$ and $Z$ are degenerate to within $ 0.2\%$.

%%%%%%%
\subsubsection*{Lepton universality in $ Z$ decays}
\label{sec:LFU:Z}
%%%%%%%

The mixing between $X$ and $Z$ also results in non-universal $Z$ boson couplings to the SM leptons, Eq.~\eqref{eq:tildeJZ}, which were measured to high precision at LEP~\cite{ALEPH:2005ab}.
We expect the strongest constraint to come from the lepton flavor universality ratio for first two generations of leptons~\cite{ALEPH:2005ab},
	\begin{equation}
	R_{\mu e} = \dfrac{\Gamma(Z\to \mu^+ \mu^-)}{\Gamma(Z\to e^+ e^-)} = \num{1.0009(28)}.
	\end{equation}
Using Eq.~\eqref{eq:tildeJZ}, we obtain
	\begin{equation}
	\delta R_{\mu e}\simeq  \dfrac{4 }{1- 4s_w^2 + 8s_w^4} \dfrac{ t_\theta g_X}{g_Z} \Big[x_{L_2} - x_{L_1} +2 s_w^2 (x_{L_1} + x_{E_1} - x_{L_2} - x_{E_2}) \Big].
	\end{equation}
for the leading new physics contributions to flavor universality ratio.
This bound is more model-dependent than the $T$ bound, since it depends on the $\U(1)_X$ charges of the leptons that change from one $\U(1)_X$ model to another. To be completely consistent, one would in principle have to perform a global electroweak fit for every $\U(1)_X$ model. However, we estimate the typical non-universal effects to be sub-leading and, thus, their inclusion would only marginally improve on the constraint obtained from the $T$ parameter in Section~\ref{sec:Tparam}.

\subsubsection*{Effective $ X$ current}
After EWSB the effective couplings of $ X $ boson (the mass eigenstate mostly composed of $X_0$) are encoded in the effective current $J_{X, {\rm eff}}$,  Eq.~\eqref{eq:tildeJX}.
In the limit
$ |s_w \hat \varepsilon_Y| < |1- r_X^2|$
the current takes a simpler form:
    \begin{equation} \label{eq:JX,eff}
    J_{X,\mathrm{eff}}^\mu  = J_X^\mu + c_\theta s_w \hat \varepsilon_Y \frac{g_Z}{g_X} \left[c_w^2 J_A^\mu + \frac{r_X^2}{1 - r_X^2} J_Z^\mu \right].   \\
    \end{equation}
For small masses, $m_X\ll m_Z$, the last term is power suppressed and we find agreement with the low-energy description of $ X $ mixing with the photon, given in Section~\ref{sec:models}. A comparison with Eq.~\eqref{eq:low-energy_charges} yields
    \begin{equation}
    \varepsilon = c_w \, \varepsilon_Y .
    \end{equation}

\subsection{Neutrino trident production}
\label{sec:NTP}

The $\Delta a_\mu$ anomaly points to a vector boson in the mass range $10$\,MeV\,$\lesssim m_X \lesssim 1$\,TeV. The viable mass window is set by cosmology (lower limit, see Section \ref{sec:cosmo}) and perturbative unitarity (upper limit, see Section \ref{sec:g-2}). An efficient complementary constraint, which cuts significantly into this parameter range, is due to limits on nonstandard neutrino trident production, i.e., the scattering of a muon neutrino on a nucleus, producing a pair of charged muons,  $\nu_\mu N \to \nu_\mu N \mu^+ \mu^-$. In combination with the constraints from the electroweak $T$ parameter, Section \ref{sec:Tparam}, it limits the $X$ mass to  $m_X \lesssim 4$\,GeV,  as we show below.

Neutrino induced production of a $\mu^+ \mu^-$ pair in the Coulomb field of a heavy nucleus is a rare electroweak process in the SM. In $\U(1)_X$ models, there is an additional tree-level contribution from the diagram with an $X_\mu$ gauge boson exchanged between $\nu_\mu$ and $\mu$ legs. The strongest bound on this contribution is due to the CCFR experiment~\cite{Mishra:1991bv}, which reported the measurement
\begin{equation}
\label{eq:CCFR}
    \frac{\sigma_\sscript{CCFR}}{\sigma_\sscript{SM}} = 0.82\pm 0.28~.
\end{equation}
for a trident cross section normalized to the SM prediction.\footnote{The CHARM-II~\cite{CHARM-II:1990dvf} and the NuTeV~\cite{NuTeV:1999wlw} bounds are weaker; however, the NuTeV experiment identified an additional background not included in the CCFR analysis, raising some concerns that the errors quoted in Eq.~\eqref{eq:CCFR} may be underestimated.}
This imposes constraints on the vector and axial vector couplings of $X_\mu$ to muons. We derive the resulting bounds on the gauge coupling $g_X$ as a function of $m_X$ for the $\U(1)_X$ models using the public code of Ref.~\cite{Altmannshofer:2019zhy} (further details can be found in Ref.~\cite{Greljo:2021npi}).
In the EFT region, $ m_X \gtrsim \SI{1}{GeV}$, this bound is approximated by~\cite{Altmannshofer:2019zhy}
    \begin{equation}\label{eq:trident_EFT}
    \frac{\sigma_\sscript{CCFR}}{\sigma_\sscript{SM}} \simeq \frac{\big( \tfrac{1}{2} + 2 s_w^2 + C_V \big)^2 + 1.13 \big(\tfrac{1}{2} + C_A\big)^2}{\big( \tfrac{1}{2} + 2 s_w^2 \big)^2 + 1.13/4},
    \end{equation}
where
    \begin{equation}
    C_{V,A} = \frac{g_X^2}{\sqrt{2} G_F m_X^2} q_\nu q_{V,A}
    \end{equation}
are the normalized NP coefficients of the effective 4-fermion interactions between muon neutrinos and vector and axial muons, respectively.
$G_F$ is the Fermi constant, controlling the overall strength of the SM contribution, while $q_{V,A}$ and $q_\nu$ are the effective $\U(1)_X$ vector and axial charges of muons and the charge of muon neutrinos, respectively (cf. Eq.~\eqref{eq:L:amu}):
    \beq
    \label{eq:trident:L}
    \L \supset
    g_X\, \overline{\mu}\, \slashed X (q_{V} - q_{A} \gamma_5) \mu + g_X\, \overline{\nu}_\mu\, q_{\nu} \slashed X P_L \nu_\mu.
    \eeq

%%%%%%%%%%%%%%%%%%%%%%%%%%%%%%%%%%%
\subsection{Neutrino oscillations and coherent elastic neutrino-nucleus scattering}
\label{sec:NSI}

Beyond the trident production, discussed in Section \ref{sec:NTP}, the $\U(1)_X$ solutions to the $(g-2)_\mu$ anomaly also lead to two other types of nonstandard neutrino interactions (NSI): the modified matter effects in neutrino oscillations and the additional contributions to coherent elastic neutrino--nucleus scattering.

The matter effects in neutrino oscillations are given by the forward scattering amplitude, i.e., at zero momentum transfer. Accordingly, the NSI contributions are well described by the EFT Lagrangian ($ f=\{e, p, n\} $)~\cite{Wolfenstein:1977ue,Mikheyev:1985zog,Antusch:2008tz,Coloma:2020gfv},
\begin{equation} \label{eq:eff_neutrino_matter_int}
    \mathcal{L}_\sscript{NSI} = -{2\sqrt{2}} {G_F} \sum_{f,\alpha\beta} \varepsilon_{\alpha \beta}^f (\overline{f} \gamma_\mu f) (\overline{\nu}_\alpha P_\LL \nu_\beta)~,
\end{equation}
even for $m_X$ well below the mass window of interest, $m_X\gtrsim 10$ MeV. In our setup, the NSI are generated at tree level by integrating out the gauge field $X_\mu$.
We use the results of a global EFT fit to the neutrino oscillations data~\cite{Esteban:2018ppq,Coloma:2020gfv,Heeck:2018nzc} to set constraints on $X_\mu$ couplings to $u$ and $d$ quarks and to electrons (cf. App.~\ref{app:NSI}).
The constraints are numerically important for the \gmu compatible parameter space. For instance, the upper limit on $g_X / m_X$ from neutrino oscillations rules out $\U(1)_{B-3 L_\mu}$ as a possible solution to the \gmu anomaly, despite relatively small couplings to quarks, $ x_q/x_\mu\ll 1$ (see Fig.~3 in Ref.~\cite{Greljo:2021npi}).

The bounds on NSI  from neutrino oscillations, Eq.~\eqref{eq:eff_neutrino_matter_int}, are sensitive to the average $\U(1)_X$ charge of the material the neutrinos propagate through, either the Earth's mantel and core or the sun.
Among other implications, this means that the value of the kinetic mixing $\varepsilon$ has no effect on the neutrino oscillation bounds, since the matter is electrically neutral.
The bounds from neutrino oscillations are relaxed for a particular set of $\U(1)_X$ models, such as $B-2L_e-L_\tau$ and $B-2L_e-L_\mu$, for which the average $\U(1)_X$ charge of normal matter is almost zero
(nuclei have, on average, roughly as many protons as neutrons, i.e., $B\simeq 2 L_e$ for normal matter).

A complementary set of constraints on NSI is due to coherent elastic neutrino--nucleus scattering~\cite{Freedman:1973yd,Drukier:1984vhf}, which was observed by the COHERENT experiment~\cite{COHERENT:2017ipa}. In this case, the EFT description in Eq.~\eqref{eq:eff_neutrino_matter_int} is not valid for the full range of $X$ masses of interest to our analysis, $10$\,MeV\,$\lesssim m_X \lesssim 4$\,GeV. Following Ref.~\cite{Greljo:2021npi}, we instead keep $X_\mu$ as a dynamical field when setting the
bounds on $g_X$, using the likelihood from Ref.~\cite{Denton:2020hop}.

%%%%%%%%%%%%
\subsection{Elastic neutrino-electron scattering}
\label{sec:other}
%%%%%%%%%%%%

The Borexino experiment measured the cross section for elastic scattering of solar neutrinos on electrons~\cite{Bellini:2011rx,Borexino:2017rsf}, which constrains possible new $X_\mu$ mediated interactions between neutrinos and electrons. The strongest bound on light vector boson interactions is obtained from $^7\mathrm{Be} $ neutrinos, which have an energy of \SI{862}{keV}.
The direct $ X $ coupling to electrons can significantly change the scattering cross section of solar neutrinos, especially due to many $\U(1)_X$ models having large couplings to muon and tau neutrinos.  The $\nu_\mu$ and $\nu_\tau$ neutrino flavors are present in the solar flux on Earth, since the initial $\nu_e$ neutrinos oscillate during propagation from the Sun. We follow the analysis of Ref.~\cite{Altmannshofer:2019zhy} to place bounds on $ g_X $ while requiring that the scattering cross section remains within $ 2 \sigma $ of the measurement.

The reactor experiments TEXONO~\cite{TEXONO:2009knm} and GEMMA~\cite{Beda:2009kx} measured the related cross section for elastic scattering of electron anti-neutrinos on electrons, while the high-energy beam experiment CHARM-II at CERN measured the cross sections for elastic $\nu_\mu e^-$ and $\bar \nu_\mu e^-$ scattering~\cite{CHARM-II:1993phx,CHARM-II:1994dzw}. These bounds can be as relevant as Borexino and will be discussed in the context of the chiral $\tilde{L}_{\mu-\tau}$ model in Section~\ref{sec:model-II}, where the electron coupling is purely axial avoiding bounds from NSI oscillations.

%%%%%%%%%%%%
\subsection{Resonance searches}
\label{sec:res}
%%%%%%%%%%%%

Several intensity-frontier collider experiments~\cite{Ilten:2018crw,Bauer:2018onh} have directly searched for a production of a vector resonance $X$ in the mass range of interest, $10\,\text{MeV}\lesssim m_X\lesssim 4\,\text{GeV}$.

In a fixed target experiment such as NA64~\cite{Banerjee:2019pds}, the vector boson is produced via bremsstrahlung process $ e N \to e N X $, where $N$ is a nucleus. The most relevant decay mode is $ X \to \mathrm{invisible} $, which is typically dominant below the dimuon threshold. The events are reconstructed from the missing energy measurements. The NA64 search~\cite{Banerjee:2019pds} constrains the $X$ couplings to electrons, which enter the prediction for the $X$ production rates. Future NA64$\mu$~\cite{Gninenko:2018ter} and $M^3$~\cite{Kahn:2018cqs} experiments will feature muons in the incoming beam and will have the potential to cover the entire parameter space of the $L_\mu - L_\tau$ model, see Fig.~2 in Ref.~\cite{Greljo:2021npi}. The fixed target experiments are effective only for $m_X \lesssim 1$\,GeV.

Another important set of direct searches was performed at $B$-factories and probed $X$ masses up to $10$\,GeV. The BaBar search for $e^+ e^- \to \mu^+\mu^- X$ in the $4\mu$ final state~\cite{BaBar:2016sci} (labeled in figures as BaBar 2016) sets fairly stringent constraints above the dimuon threshold. BaBar also searched for a radiative return process $e^+ e^- \to \gamma X$ with $X$ decaying to $e^+ e^-$ or $\mu^+ \mu^-$~\cite{BaBar:2014zli} (BaBar 2014 and LHCb) or to invisible~\cite{BaBar:2017tiz} (BaBar 2017).
The LHC searches extend the exclusion to even larger $X$ masses, such as the LHCb search for $X\to \mu^+\mu^-$~\cite{LHCb:2017trq,LHCb:2019vmc} and the CMS search for $Z \to \mu^+ \mu^- X \to 4 \mu$~\cite{CMS:2018yxg}.

In all the aforementioned searches, the $X$ decays are prompt in the parameter space relevant for \gmu. We use the \texttt{DarkCast} code~\cite{Ilten:2018crw}, which comes with the compilation of relevant bounds, to set limits on the gauge coupling $g_X$ as a function of the mass $m_X$. Crucially, the above bounds are model dependent; for instance, the constraints from dimuon resonance searches could be removed by introducing additional invisible $X$ decays to a light dark sector.

\section{Phenomenological implications}
\label{sec:pheno:implications}

We explore next the phenomenological implications of experimental constraints on minimal anomaly-free $\U(1)_X$ models as candidates for explaining the $(g-2)_\mu$ anomaly. In Section~\ref{sec:generic}, we show that a combination of trident and electroweak precision tests, assuming the SM is the only source of EWSB, i.e., that $\U(1)_X$ is broken by SM singlet scalar(s), leads to the upper bound $m_X\lesssim 4$~GeV. In Section~\ref{sec:global:constr}, we include other experimental constraints that are particularly relevant for this low $X$ mass region and perform a global analysis of the complete set of 419 phenomenologically distinct vector-like $\U(1)_X$  models with charges up to 10. The discussion of chiral models is relegated to Section \ref{sec:model-II}.

\subsection{Generic upper bound on $m_X$}
\label{sec:generic}
Ref.~\cite{Altmannshofer:2014cfa} demonstrated that the trident production sets stringent constraints on the viable parameter space of the \gmu solution in the \Umt model, limiting
    \begin{equation}
    \label{eq:mX:nokin}
    m_X \lesssim 0.5\,\text{GeV}\qquad \text{(trident)}.
    \end{equation}
The above constraint was shown in Ref.~\cite{Greljo:2021npi} to hold for all the $ \U(1)_X $ models in the limit of vanishingly small kinetic mixing ($ \varepsilon =0 $), where the muon couplings to $X_\mu$ are due to muon $\U(1)_X$ charges. This was achieved by marginalizing the trident constraint over all values of $x_{L_2}$ and $x_{E_2}$---the $\U(1)_X$ charges of the second generation left-handed lepton doublet and right-handed lepton singlet, respectively---with the results of the analysis shown in Fig.~1 of Ref.~\cite{Greljo:2021npi}.
In the $ \varepsilon =0 $ limit, the vector and axial couplings of $X_\mu$ to muons are directly correlated to the $\U(1)_X$ charges of left- and right-handed muons, with the vector charge given by $(x_{L_2} + x_{E_2})/2$ and the axial charge by $(x_{L_2} - x_{E_2})/2$.
This is the reason why the trident bound so efficiently constrains the possible $ \U(1)_X $ solutions of \gmu.
Since \gmu  requires near-vectorial couplings, that is, comparable $ x_{L_2} $ and $ x_{E_2} $, a considerable $x_{L_2}$ is required.\footnote{In the renormalizable, anomaly-free  $\U(1)_X$ vector models, the $X_\mu$ couplings to charged leptons are diagonal in the mass basis and do not lead to cLFV.
The new physics contribution to \gmu is due to a muon and an $X_\mu$ running in the loop. The bottom-up phenomenological vector model of Ref.~\cite{Altmannshofer:2016brv}, in which the new physics contribution to \gmu is due to a tau and an $X_\mu$ running in the loop, is difficult to UV complete in a gauged $\U(1)_X$. Such a solution requires, for instance, an extended scalar sector beyond our minimal setup (see, e.g., Ref.~\cite{Cheng:2021okr}).}
This, in turn, implies similar couplings of $X_\mu$ to muons and muon neutrinos, the upper components of the $L_2$ doublet, and, thus, a sizable neutrino trident production $\nu_\mu N \to \nu_\mu N \mu^+ \mu^-$.

When $\varepsilon\ne 0$, the direct correlation between the NP shift in $(g-2)_\mu$ and the trident production no longer applies. The kinetic mixing between $ X_\mu$ and the photon in the low-energy Lagrangian~\eqref{eq:L:benchmark}
shifts the coupling of upper and lower components of the isospin doublets independently and decorrelates muon and neutrino couplings to $ X_\mu $.
An instructive example is the $B- 2L_e - L_\tau$ model solution to $(g-2)_\mu$, in which the muon couplings to $X_\mu$ are generated entirely through its mixing with the photon, while there are no couplings to muon neutrinos and, thus, essentially no trident bounds on $X_\mu$.\footnote{The typical flavor composition of the initial neutrino flux in these experiments is dominantly $\nu_\mu$, whereas the next-largest, $\nu_e$, constitutes less than a percent of the neutrino flux (see, e.g., Ref.~\cite{CHARM-II:1989srx}). Suppressing the $X $ coupling to $\nu_\mu$ is sufficient to make the CCFR bound irrelevant in the parameter region that leads to the solution of $(g-2)_\mu$.} Alternatively, for $ m_X $ comparable to $ m_Z $, one might imagine a scenario where the $ J_X $ and $ J_Z $ contributions to $ J_{X,\mathrm{eff}}$~\eqref{eq:JX,eff} conspire to a similar effect. To account for these possibilities, we include EWPT as a complementary constraint to set a model-independent upper bound on $ m_X $ for viable \gmu solutions.

\begin{figure*}[t]
	\centering
     \includegraphics[width=15cm]{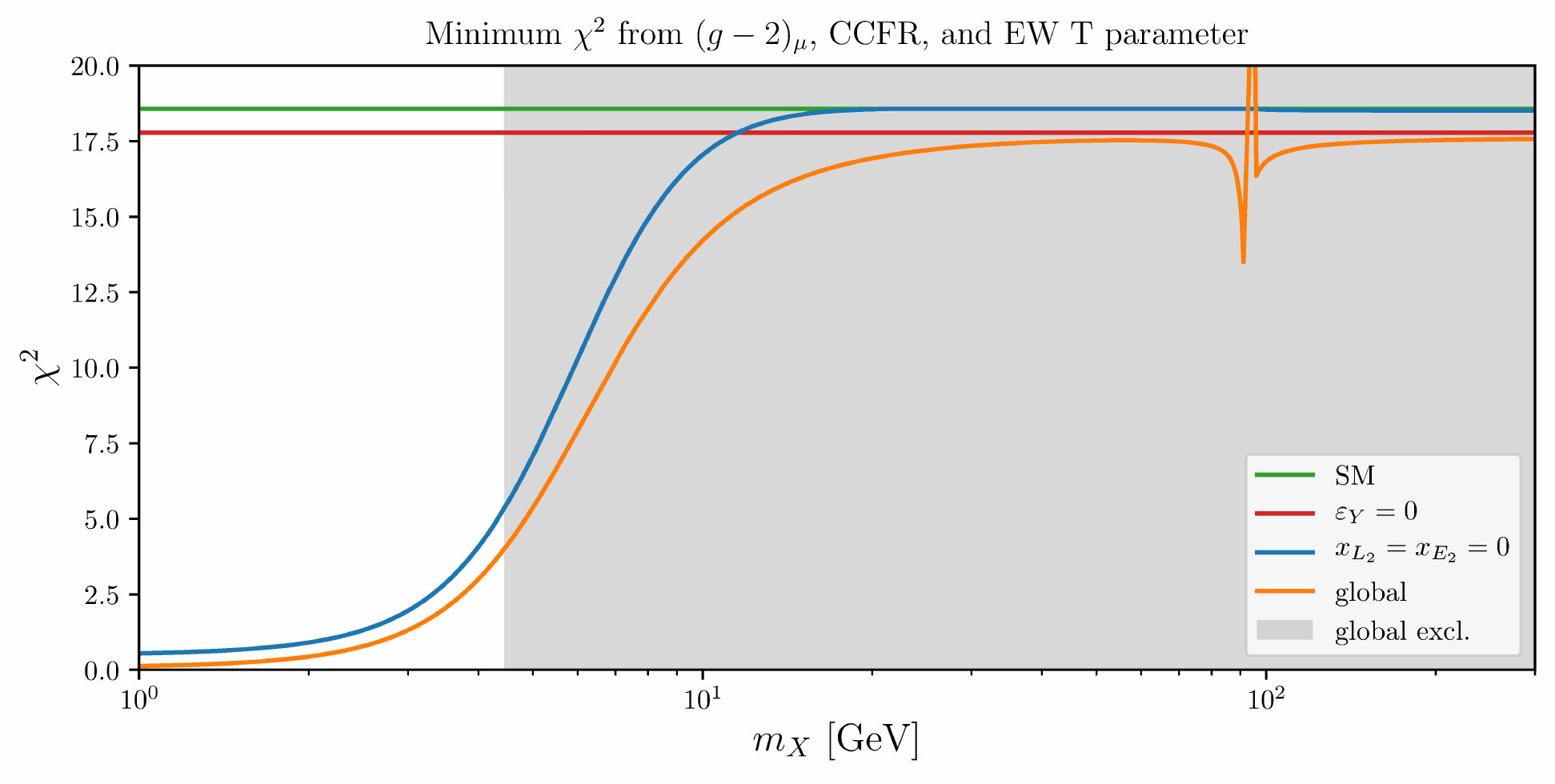}
	\caption{Minimum $\chi^2$ values in a model with a new $X_\mu$ vector boson coupling to second generation leptons with charges $x_{L_2}$ and $x_{E_2}$ and a kinetic mixing with the hypercharge gauge boson $\varepsilon_Y$ from a fit to experimental data taking into account $(g-2)_\mu$, CCFR, and the EW $ T $ parameter (see the text for details).
	The orange curve is the global best fit $ \chi^2 $, the blue is the best fit in the absence of muon charges (dark photon), the red is the best fit for vanishing kinetic mixing, and the green line is the SM $ \chi^2 $.
	In the shaded grey region, $(g-2)_\mu$ cannot be explained by any values of $x_{L_2}$, $x_{E_2}$, and $\varepsilon_Y$ while at the same time satisfying the constraints from CCFR and the electroweak $T$ parameter.} \label{fig:NTP}
\end{figure*}

Given $ m_X $, the $ X $ boson couplings to second generation leptons~\eqref{eq:JX,eff} are determined through a known combination of parameters $ g_X x_{L_2} $, $ g_X x_{E_2} $, and $ \varepsilon_Y $. These determine the $\U(1)_X$ shifts in $ T $~\eqref{eq:T_contribution}, the neutrino trident cross section~\eqref{eq:trident_EFT}, and \gmu~\eqref{eq:amu_NP}.
We use these observables to construct a combined $ \chi^2 $. For each $m_X$ in  $\SIrange{1}{300}{GeV}$, we then find the $\chi^2$ minimum by varying $ g_X x_{L_2} $, $ g_X x_{E_2} $, $ \varepsilon_Y $, shown as the orange line in Fig.~\ref{fig:NTP}, to be compared with the SM value in green. The case of no kinetic mixing, $\varepsilon_Y=0$, is shown with the red line, while the blue line shows the case where the couplings to $X_\mu$ are entirely due to kinetic mixing.

We observe that at low $m_X$ masses, the best fit is similar to the solution where the $X_\mu$ couplings are exclusively generated by the kinetic mixing. However, as the $m_X$ mass increases there is a growing tension between a large value of $\varepsilon_Y$, which would minimize the size of the $X_\mu$ couplings to neutrinos and the effect of the trident constraints, and a small value as preferred by EWPT constraints.
The net effect is that a combination of trident production and EWPT constraints limit the $X$ boson solutions to \gmu to be rather light:
    \begin{equation} \label{eq:mX_upper_bound}
    m_X \lesssim 4\,\text{GeV}\qquad \text{(trident+$T$)}.
    \end{equation}
It is worth reiterating that the only assumptions entering  this bound are that the $\U(1)_X$ model is renormalizable.
Higher-dimensional operators could be added to the Lagrangian~\eqref{eq:lag_EWX} to modify the bound; however, this goes beyond the minimality assumed in this paper.

\subsection{Global constraints on anomaly-free vector-like $\U(1)_X$ models}
\label{sec:global:constr}

Having set an overall upper bound on $ m_X $, we perform a scan over all vector-like models in order to determine those that can account for the \gmu discrepancy and at the same time satisfy all the constraints discussed in Section~\ref{sec:constraints}. We assume that the sterile neutrinos are heavy enough so as not to affect the low energy phenomenology.
The $\U(1)_X$ charges in the vector-like models then constitute a 3-parameter family of possible charge assignments, with the charges $x_f$ for each SM fermion given by
\begin{equation}\label{eq:X_VL}
    x_f=c_B B - c_e L_e - c_\mu L_\mu - c_\tau L_\tau, \quad \text{where}\quad c_B= \frac{c_e + c_\mu + c_\tau}{3},
\end{equation}
with $B$, $L_{e,\mu,\tau}$ the baryon and individual lepton numbers and $c_{e, \mu, \tau}$ continuous parameters. For the 419 vector-like models with $|x_f|\leq 10$, these take specific rational values that can be found in Ref.~\cite{Greljo:2021npi}.
An overall factor can be absorbed into the gauge coupling $ g_X $, which reduces the number of independent physical parameters to two.
Since it is well-known that the $ L_\mu - L_\tau $ model can explain \gmu while passing all constraints, we express the two-dimensional model space in Eq.~\eqref{eq:X_VL} in a way that makes it apparent how closely a given model resembles the $ L_\mu - L_\tau $ charge assignments,
\begin{equation}\label{eq:X_VL_polar}
    x_f \propto \sin(\alpha)\big(L_e-L_\mu)+\cos(\alpha)\big(B/3-L_\mu\big)+ R\big(L_\mu-L_\tau\big).
\end{equation}
The polar angle $\alpha$ determines the ratio of baryon and electron charges, while the radial parameter $R$ gives the ``closeness'' to $ L_\mu - L_\tau $, with $ R \to \infty $ corresponding to the exact $ L_\mu - L_\tau $ limit.
Note that $L_e$ and $B$, with coefficients proportional to $\sin(\alpha)$ and $\cos(\alpha)$, respectively, enter the parameterization in a combination with $L_\mu$ in such a way that models given by Eq.~\eqref{eq:X_VL_polar} are anomaly-free for arbitrary values of $\alpha$ and $R$.

We use the parametrization \eqref{eq:X_VL_polar} to systematically treat the experimental constraints on the vector-like models. Since only low $X$ mass explanations of \gmu are still possible, cf. Eq.~\eqref{eq:mX_upper_bound}, we can focus on the low-energy Lagrangian~\eqref{eq:L:benchmark}.
For each vector-like model, the physics is determined by the $ X $ mass $ m_X$, the gauge coupling $ g_X $, and the kinetic mixing parameter $ \varepsilon $.
To reduce the complexity of the analysis, we identify $ m_X \simeq \SI{200}{MeV} $ as the mass with the best possibility of explaining \gmu: for masses above the di-muon threshold the resonance searches at LHCb and Babar provide very strong constraints.
All the remaining important bounds, in particular the bounds on NSI from neutrino oscillations, Borexino, NA64, and COHERENT, limit the $X$ mass from below, and thus taking $m_X$ close to the di-muon threshold minimizes their importance.

\begin{figure*}[t]
	\centering
     \includegraphics[width=\textwidth]{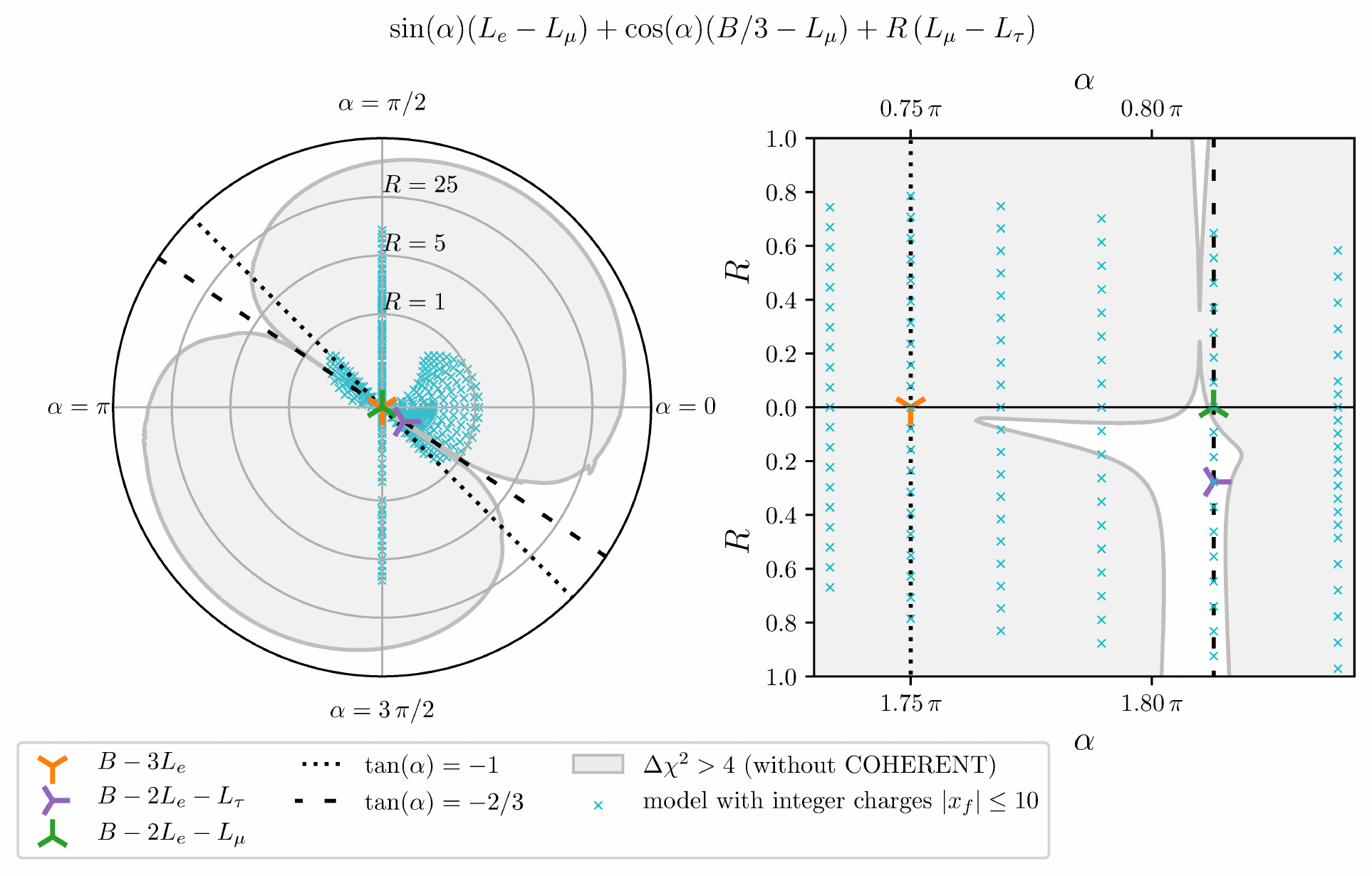}
	\caption{
	Bounds on vector-like gauge model solutions to \gmu, parameterized by Eq.~\eqref{eq:X_VL_polar}, at $ m_X = \SI{200}{MeV} $.
	Inside the grey region, an explanation of \gmu is excluded by the combination of data from neutrino oscillations in matter, NA64, Borexino, BaBar 2014 (see text for details).
	The blue crosses correspond to models with integer charges $|x_f|\leq10$.
	Linear and logarithmic scales are used for $R<1$ and $R>1$, respectively.
	} \label{fig:NSI:circle}
\end{figure*}

To narrow down the possible candidates for explaining \gmu, we scan over $\alpha$ and $R$ in Eq.~\eqref{eq:X_VL_polar}.
For each point in the $(\alpha, R)$ model space, we construct a $\Delta \chi^2 $ function that takes into account the bounds from $ \Delta a_\mu $, Borexino, and NSI~osc. and minimize it with respect to $ (g_X, \, \varepsilon) $ at $ m_X = \SI{200}{MeV} $.
As an additional constraint, we require $ (g_X, \, \varepsilon) $ to be in the range allowed by NA64 and BaBar2014 searches, as implemented in \texttt{DarkCast}.
We deem a model to be excluded as a viable explanation of the $(g-2)_\mu$ anomaly, if the minimum $\Delta \chi^2$ value is above 4. This is to be compared with the SM value for $\Delta \chi^2$, which is $\Delta \chi^2\big|_{\rm SM}=18.3$), i.e., we discard models that are in tension with either \gmu or any individual constraint by more than 2$\sigma$. Further details on the construction of $\Delta \chi^2$ are given in Appendix \ref{app:global:chi2}.

In the left panel of Fig.~\ref{fig:NSI:circle}, we show the entire excluded region in the model space.
For illustration, we mark all the models with integer charges $|x_f|\leq10$ with small blue crosses and the three benchmark models $B-3 L_e$, $B-2 L_e -L_\tau$, and $B-2 Le -L_\mu$ by orange, purple, and green tripods.
As expected, for large values of the radial parameter, $R\gtrsim 50$, all the models are viable, irrespective of the value of $\alpha$. All such  models closely resembles the $L_\mu-L_\tau$ model with only small deviations.

For smaller $R$ values the viability of the models is strongly $\alpha$ dependent, i.e., dependent on the baryon-to-electron charge ratio.
For small values of $R$, the viable models are restricted to a narrow region around $\tan(\alpha)= -2/3$ (depicted as the black dashed line in Fig.~\ref{fig:NSI:circle} left) so that the $\U(1)_X$ electron charge is roughly $-2$~times the baryon charge. For such charge assignments, normal planetary matter, which has on average about as many protons as neutrons, is almost neutral under $\U(1)_X$. As discussed in Section~\ref{sec:NSI}, this results in the reduced relevance of neutrino oscillation constraints.
By contrast, the $\U(1)_X$ models in which normal matter carries a considerable $\U(1)_X$ charge are ruled out as an explanation of \gmu precisely because of the excessive contributions to the NSI.
There are only two ways for a model to both explain \gmu and predict the atoms of the most common elements to be essentially $\U(1)_X$-neutral.
Either $\tan(\alpha)\approx -2/3$ so the electron and nucleon charges approximately cancel, or $R\gg1$ so the electron and nucleon charges are small compared to the muon charge.
The only model in which all atoms are exactly $\U(1)_X$-neutral is the $L_\mu-L_\tau$ model, i.e.,\ $ R \to \infty $.
For models that differ  considerably from $L_\mu-L_\tau$, i.e., for \ $R\ll 1$, the cancellation of $\U(1)_X$ charges in normal matter is the only way to avoid the stringent NSI bounds.

In the right panel of Fig.~\ref{fig:NSI:circle}, we show the region around $\tan(\alpha)\approx -2/3$ for $R<1$. Again, we mark the models with integer charges $|x_f|\leq10$ with small blue crosses and the three sample models $B-3 L_e$, $B-2 Le -L_\tau$, and $B-2 Le -L_\mu$ with orange, purple, and green tripods, respectively.
In the lower half of the plot $\alpha$ is shifted by $\pi$ compared to the upper half of the plot, and $R$ is always positive.\footnote{%
We can restrict ourselves to positive values of $R$, since taking $R\to-R$ in Eq.~\eqref{eq:X_VL_polar} is equivalent to taking $\alpha\to \alpha+\pi$.
}
We observe that for small values of $R$, the only models that can pass the neutrino oscillation bounds either lie in a very narrow region close to $\tan(\alpha)= -2/3$, where nucleon and electron $\U(1)_X$ charges cancel inside atoms, or at around $R\approx 0.1$ in the region $-1\lesssim \tan(\alpha)\lesssim -2/3$.

\begin{table}[t]
\setlength{\tabcolsep}{.7em}
\vspace{0.2cm}
\begin{center}
\begin{tabular}{rrrrrrrrrc}
\hline\hline
\multicolumn{4}{c}{Charges} & \multicolumn{5}{c}{Best-fit values for $m_X=200$ MeV} & Fig. \\
$c_{B}$ & $c_{e}$ & $c_{\mu}$ & $c_{\tau}$ & ~$\varepsilon\,{e}/{g_X}$ && $g_X$~~~~~~ && $\Delta \chi^2$ & panel
\\
\hline
$0$ & $0$ & $1$ & $-1$ & $0.01$ && $1.44 \times 10^{-3}$ && $0.41$& Fig. \ref{fig:Lmu-Ltau}
\\
$3$ & $-7$ & $-1$ & $-1$ & $-6.97$ && $2.17 \times 10^{-4}$ && $2.00$& Fig. \ref{fig:viable_models} f)\\
$3$ & $-6$ & $7$ & $-10$ & $-6.26$ && $1.02 \times 10^{-4}$ && $2.60$& Fig. \ref{fig:viable_models} e)\\
$3$ & $-6$ & $6$ & $-9$ & $-6.21$ && $9.90 \times 10^{-4}$ && $2.84$& Fig. \ref{fig:viable_models} d)\\
$3$ & $-6$ & $5$ & $-8$ & $-6.16$ && $1.19 \times 10^{-4}$ && $3.13$& Fig. \ref{fig:viable_models} c)\\
$3$ & $-6$ & $4$ & $-7$ & $-6.11$ && $1.31 \times 10^{-4}$ && $3.47$& Fig. \ref{fig:viable_models} b)\\
$3$ & $-6$ & $3$ & $-6$ & $-6.08$ && $1.30 \times 10^{-3}$ && $3.90$& Fig. \ref{fig:viable_models} a)\\
\hline\hline
\end{tabular}
\end{center}
\caption{The $\U(1)_X$ charges parametrized by $c_{B,e,\mu, \tau}$ in \eqref{eq:X_VL} (first four columns),  and the best-fit values of the mixing parameter (fifth column) and gauge coupling (sixth column), as well as the $\chi^2$ value at the minimum taking $m_X=200$ MeV, for vector-like $\U(1)_X$ models that can explain the measurement of $(g-2)_\mu$ within $2\sigma$ and pass the bounds from NSI osc., COHERENT, Borexino, NA64, and BaBar.}
\label{tab:viable_models}
\end{table}

The constraints from neutrino oscillations on NSI already exclude most of the parameter space for $R\lesssim 10$.
Another potentially relevant constraint in this region of model space is due to coherent elastic neutrino-nucleus scattering as measured by the COHERENT experiment.
We impose this bound on all the models with integer charges $|x_f|\leq10$ that are not already excluded by the neutrino oscillation data. (Since deriving the COHERENT bound is computationally expensive this constraints was not include in Fig.~\ref{fig:NSI:circle}.)
All the  models that are not yet excluded before applying the COHERENT constraint appear in the right panel of Fig.~\ref{fig:NSI:circle} as small blue crosses on white background, apart from the $L_\mu-L_\tau$ model, which is not shown, since it corresponds to $R\to\infty$ limit.
For all these models, we add the contribution from COHERENT to the previous $\Delta \chi^2$ function and minimize the newly obtained $\Delta \chi^2$ function at $m_X=\SI{200}{MeV}$ with respect to $ (g_X, \, \varepsilon) $.
The charges and best-fit values of the models with $\Delta \chi^2<4$ are shown in Table~\ref{tab:viable_models}.
This table shows models belonging to three distinct classes, which we discuss separately next.

\begin{figure*}[t]
	\centering
     \includegraphics[width=\textwidth]{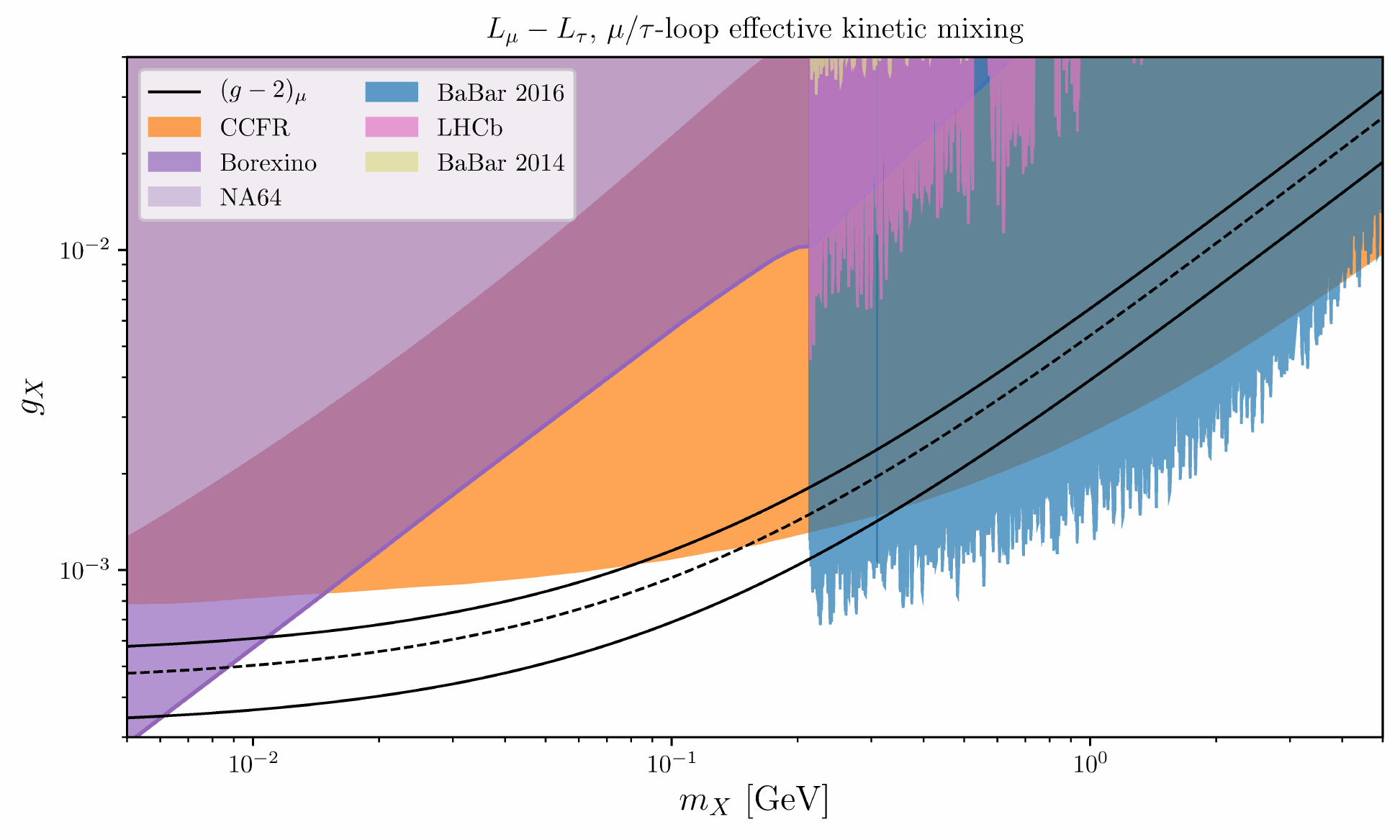}
	\caption{
	Parameter space of the $L_\mu-L_\tau$ model with the parameter space that explains the central value of $\Delta a_\mu$ denoted with dashed black line (the $2\sigma$ band by solid black lines). The exclusions by different measurements are color coded as given in the legend.
	} \label{fig:Lmu-Ltau}
\end{figure*}

\subsubsection*{The $L_\mu-L_\tau$ model}
 The $L_\mu-L_\tau$ model has the lowest $\Delta \chi^2$ value.
 It is able to explain \gmu comfortably without causing tensions with other experimental constraints.
 The best-fit point is found for vanishing kinetic mixing $\varepsilon$, however even for kinetic mixing values that one would typically obtain if this were generated at one loop, $\varepsilon\sim g_X\, e/(16\pi^2)$, the $L_\mu-L_\tau$ model can still explain $(g-2)_\mu$ without being excluded by other constraints. In  Fig.~\ref{fig:Lmu-Ltau} we show an example where the kinetic mixing is assumed to be induced by the muon and tau running in the loop.

\subsubsection*{The class $3 B-6 L_e -3 L_\tau +N (L_\mu-L_\tau)$ with $N\in\{3,4,5,6,7\}$}
\begin{figure}[t]
	\centering
     \includegraphics[width=\textwidth]{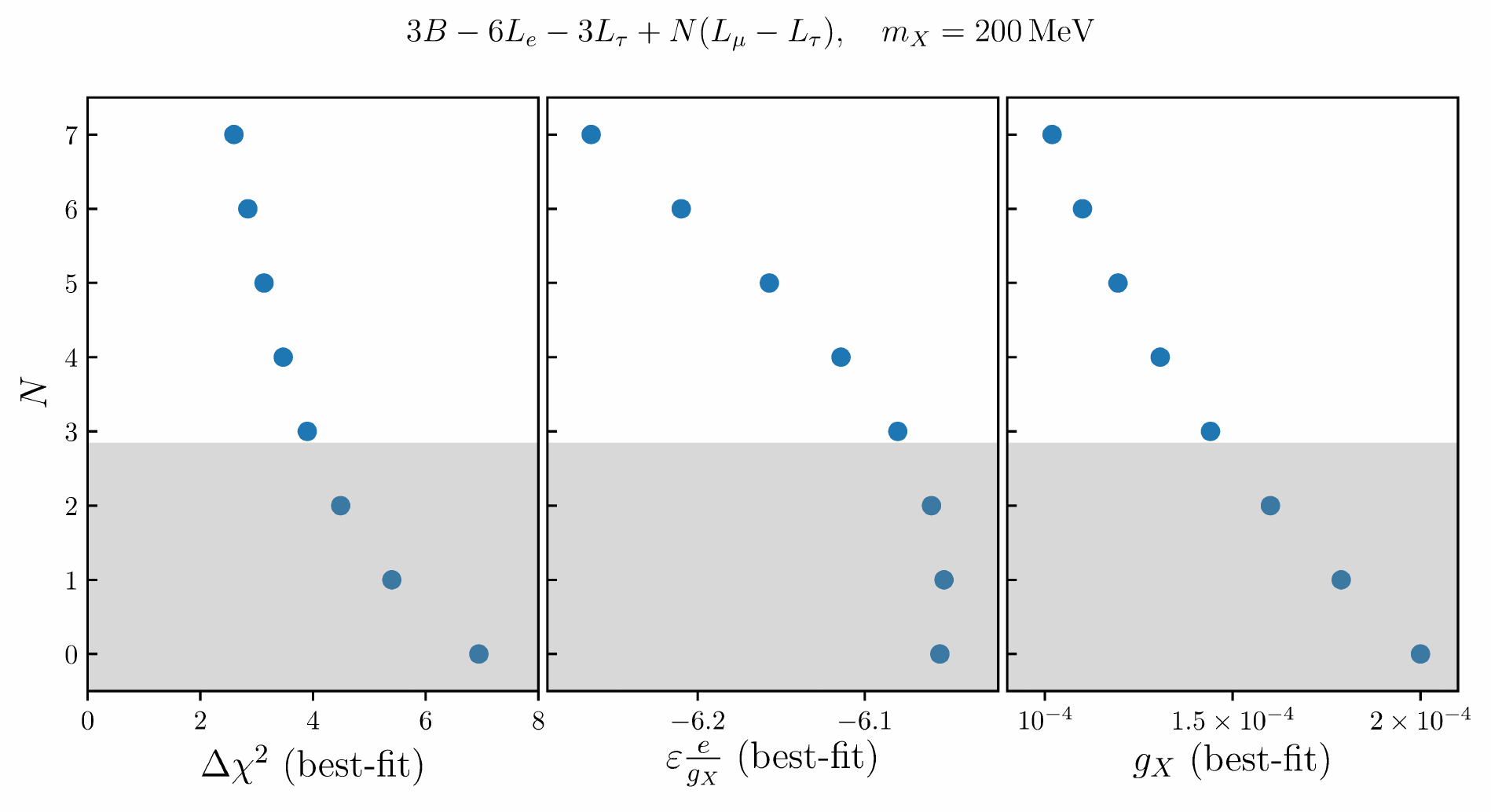}
	\caption{The comparison of minimal $\Delta \chi^2$ values (left panel), as well as best-fit values for the kinetic mixing parameter (middle) and the gauge coupling (right) for the vector-like $\U(1)_X$ models with the SM fermion charges parametrized as $x_f=3 B-6 L_e -3 L_\tau +N (L_\mu-L_\tau)$, with $N\in\{0,1,2,3,4,5,6,7\}$.
	The $\Delta \chi^2$ function takes into account the measurement of $(g-2)_\mu$ and the bounds from NSI osc., COHERENT, Borexino, NA64, and BaBar while the $X$ mass was set to 200 MeV.
	The grey shaded region corresponds to $\Delta \chi^2>4$.
	The five models with $\Delta \chi^2<4$ are also listed in Table~\ref{tab:viable_models}.
	} \label{fig:model_comparison}
\end{figure}

These models satisfy $\tan(\alpha)=-2/3$, i.e.,\ they lie on the dashed vertical line in Fig.~\ref{fig:NSI:circle} and are all within the narrow region of model space where the $\U(1)_X$ charges of electrons and nucleons in atoms approximately cancel.
In Fig.~\ref{fig:NSI:circle}, only the models with $N<-2$ were excluded. The addition of the COHERENT constraint now leads to the exclusion of all models with $N<3$.
As shown in Table~\ref{tab:viable_models}, larger values of $N$ generically correspond to lower $\Delta \chi^2$ values, that is, they can satisfy the constraints more easily.
This is visualized in the left panel of Fig.~\ref{fig:model_comparison}, where we show the $\Delta \chi^2$ values for $N\geq0$ with the grey shaded region corresponding to $\Delta \chi^2>4$.
The reason for the decrease in $\Delta \chi^2$ as $N$ increases is that the muon charge increases as well and thus a smaller gauge coupling is requires to explain \gmu.
This can be seen in the sixth column of Table~\ref{tab:viable_models} and in the right panel of Fig.~\ref{fig:model_comparison}.
Since an increase in $N$ does not affect electron and baryon charges, the smaller values of the gauge coupling lead to weaker constraints from NSI.
Ultimately, the models become more similar to the $L_\mu-L_\tau$ model as $N$ increases.

\begin{figure}[t]
	\centering
    \includegraphics[width=\textwidth]{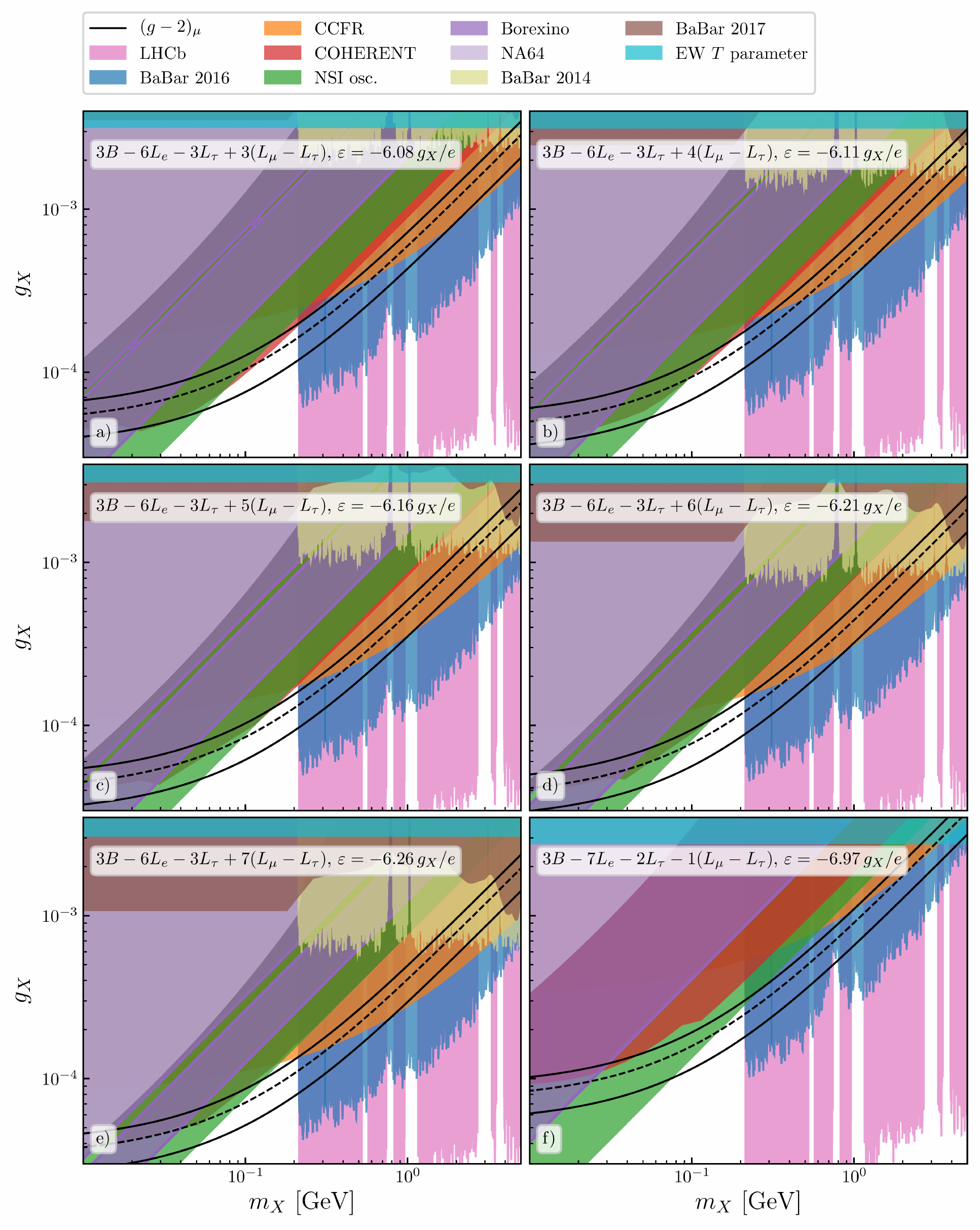}
    %\vspace{-2em}
	\caption{Parameter space of viable vectorlike $\U(1)_X$ models that can explain \gmu (the  parameter space for $L_\mu-L_\tau$ is shown in Fig. \ref{fig:Lmu-Ltau}).
	The central value of $\Delta a_\mu$ is indicated by a dashed black line (the $2\sigma$ band by solid black lines).
    The exclusions by different measurements are color coded as given in the legend.
	} \label{fig:viable_models}
\end{figure}

At the best-fit point, not only the $\Delta \chi^2$ value and the gauge coupling $g_X$ but also the kinetic mixing $\varepsilon$ is correlated with $N$, which can be understood as follows:
The COHERENT bound is due to a measurement of the  coherent elastic neutrino-nucleus scattering in cesium iodide (CsI). Accordingly, the COHERENT bound becomes less constraining if the $\U(1)_X$ charges of protons and neutrons in CsI approximately cancel.
Cesium and iodine have neutron-to-proton ratios $78/55$ and $74/53$, respectively, such that the average neutron-to-proton ratio in CsI is $1.4$ to a very good approximation.
In the $3 B-6 L_e -3 L_\tau +N (L_\mu-L_\tau)$ class of models, the effective $\U(1)_X$ charge of neutron and proton (taking into account the kinetic mixing $\varepsilon$) is $3$ and $3+\varepsilon\,{e}/{g_X}$, respectively. CsI therefore becomes approximately $\U(1)_X$-neutral for
\begin{equation}\label{eq:vanising_COHERENT}
 3+\varepsilon\frac{e}{g_X} \approx -(1.4\times 3)
 \qquad\implies \qquad
 \varepsilon\frac{e}{g_X} \approx - 7.2\,.
\end{equation}
The Borexino bound, on the other hand,  is due to a measurement of elastic neutrino-electron scattering and vanishes if the effective $\U(1)_X$ charge of the electron is zero, i.e., when
\begin{equation}\label{eq:vanising_Borexino}
 -6 - \varepsilon\frac{e}{g_X} = 0
 \qquad\implies \qquad
 \varepsilon\frac{e}{g_X} = -6\,.
\end{equation}
For small $N$, the relatively large gauge coupling needed to explain \gmu makes the Borexino bound very sensitive to deviations from Eq.~\eqref{eq:vanising_Borexino}.\footnote{%
An exception is the model space close to $N=0$, where the $\U(1)_X$-coupling of the muon neutrino vanishes, which weakens the Borexino bound.
}
As $N$ increases and the gauge coupling decreases, larger deviations from Eq.~\eqref{eq:vanising_Borexino} are allowed by Borexino and the best fit value of $\varepsilon\,{e}/{g_X}$ moves closer to Eq.~\eqref{eq:vanising_COHERENT} in order to better satisfy the COHERENT bound.
This also means that the COHERENT bound is considerably more constraining at the best fit point for smaller $N$, which in turn leads to a larger $\Delta \chi^2$ value.

The bounds on $3 B-6 L_e -3 L_\tau +N (L_\mu-L_\tau)$ with $N\in\{3,4,5,6,7\}$ are shown in Fig.~\ref{fig:viable_models} in panels a)
through e), in the plane of the $X$ mass $m_X$ and the gauge coupling $g_X$, while for each plot the kinetic mixing $\varepsilon$ is set to its best-fit value shown in Table~\ref{tab:viable_models}.
In these plots, we can observe the interplay of Borexino and COHERENT constraints described above.
For small $N$, e.g., for $N=3$ shown in panel a), $g_X$ has to be relatively large in order to explain \gmu.  Consequently, the Borexino bound (shown in dark purple) requires $\varepsilon$ to be very close to Eq.~\eqref{eq:vanising_Borexino}. This in turn leads to a relatively strong COHERENT bound (shown in red), which then excludes part of the parameter region preferred by \gmu in the $m_X$ mass range between $\SI{100}{MeV}$ and $\SI{200}{MeV}$.
Comparing this to the models with increasing $N$ in panels b) through e), one observes that the size of $g_X$ preferred by \gmu decreases due to the increasing muon charge, the Borexino bound allows for smaller $\varepsilon$, and the COHERENT bound becomes less and less important until it becomes weaker than the bound from the neutrino oscillations for $N=7$, cf.\ panel e).

The plots in Fig.~\ref{fig:viable_models} also demonstrate that for $m_X$ above the dimuon threshold, resonance searches, in particular from BaBar and LHCb, prevent an explanation of \gmu, whereas for lower values of $m_X$, neutrino oscillation data provides the strongest bound.
This justifies the choice of $m_X=\SI{200}{MeV}$ used in Figs.~\ref{fig:NSI:circle} and~\ref{fig:model_comparison} and Table~\ref{tab:viable_models}.

It is interesting that the CCFR bound from neutrino trident production (shown in orange) becomes less constraining for smaller values of $N$ and even completely disappears for $N=0$.
The reason is that the effective $\U(1)_X$ couplings of the charged muon and muon neutrino are $N-\varepsilon\,{e}/{g_X}$ and $N$, respectively.
For sizable $\varepsilon$ and decreasing $N$, the region in parameters space where \gmu can be explained is, therefore, less and less excluded by the CCFR bound. In the large $N$ limit, on the other hand, the model is closer and closer to the $L_\mu-L_\tau$ model and the CCFR constraint becomes increasingly important.

\subsubsection*{The class $3 B-7 L_e -2 L_\tau +N (L_\mu-L_\tau)$ with $N=-1$}
These models satisfy $\tan(\alpha)=-7/9$ so that the cancellation of $\U(1)_X$ charges inside atoms is less efficient than in the models with $\tan(\alpha)=-2/3$.
Nevertheless, there is a window roughly around $N=-1$ where the bound from neutrino oscillations is relatively weak and an explanation of \gmu is possible (cf.\ right panel of Fig.~\ref{fig:NSI:circle}).
The virtue of this class of models is that both the Borexino and the COHERENT bounds can be comfortably satisfied.
This is apparent from the constraints displayed in Fig.~\ref{fig:viable_models}, panel f), which show that for $m_X$ below the dimuon threshold the neutrino oscillations provide the most relevant bound in the $(g-2)_\mu$ band, whereas COHERENT and Borexino are far less important.
While the COHERENT bound vanishes for $\varepsilon\,{e}/{g_X} \approx - 7.2$, as in Eq.~\eqref{eq:vanising_COHERENT}, the effective $\U(1)_X$ charge of the electron in the present class of models is $-7 - \varepsilon\,{e}/{g_X}$ such that the Borexino bound disappears for $\varepsilon\,{e}/{g_X} = -7$.
Consequently, values of $\varepsilon\,{e}/{g_X} \approx -7$ lead to very weak bounds for both Borexino and COHERENT. The best fit values for $\varepsilon\,{e}/{g_X}$ are indeed found to be very close to $-7$ by the global scan (see Table~\ref{tab:viable_models}).
The only model in the class $3 B-7 L_e -2 L_\tau +N (L_\mu-L_\tau)$ that enters the list of viable models in Table~\ref{tab:viable_models} is the one with $N=-1$, but it actually reaches the second lowest $\Delta \chi^2$ of all the models in this list.
This model is also notable for particularly weak bounds from neutrino trident production.

%%%%%%%%%%%%%%%%%%%%%%%%%%%%%%%%%%%
\section{Chiral models}
\label{sec:model-II}

There are 21 chiral charge assignments in the charge range $[-10,10]$ (not counting permutations) in which lepton masses cannot be fully realised at the renormalizable level~\cite{Greljo:2021xmg}. Upon careful inspection of the charge assignments in the entire class of the chiral $\U(1)_X$ models, we focus on one of the best performing models, the model $\tilde L_{\mu-\tau}$. Quite generally, for these types of models the constraints in the parameter region relevant for $(g-2)_\mu$ will be minimized, if the $\U(1)_X$ charges of the quarks vanish, and the electron coupling is purely axial, in order to avoid the NSI bounds, while the vector-like muon charge is as large as possible. The $\tilde L_{\mu-\tau}$ model satisfies all  of the above requirements. Even so,  the $\tilde L_{\mu-\tau}$ model is excluded as the explanation of the $(g-2)_\mu$ anomaly by a set of complementary constraints, as we show below, and we expect the same to be true for the other chiral models.

In the $\tilde L_{\mu-\tau}$ model, the $ \U(1)_X $ charges of the fermions are  given by~\cite{Greljo:2021npi}
\begin{subequations} \label{eq:X_Lt}
\begin{align}
    (x_{L_1},\,x_{L_2},\,x_{L_3})
&   = (-1, 7, -6),  \\
     (x_{E_1},\,x_{E_2},\,x_{E_3})
&   = (1, 6, -7),   \\
      (x_{N_1},\,x_{N_2},\,x_{N_3})
&   = (-7, -2, 9),   \\
    x_{Q_i,D_i,U_i}
&   = 0,
\end{align}
\end{subequations}
where the first line is for the left-handed  leptons, the second and the third lines are for the right-handed charged leptons and neutrinos, respectively, while the quarks do not carry a $\U(1)_X$ charge.

Due to the $\U(1)_X$ charge assignment, the charged lepton Yukawas are forbidden at the renormalizable level and only arise once the $\U(1)_{X}$ gauge symmetry is spontaneously broken.
For concreteness let us consider the minimal possibility---that it is due to a SM-singlet scalar $\phi$ with $\U(1)_{X}$ charge $x_{\phi} = 1$ that develops a VEV $\langle \phi \rangle\ne0$.
The diagonal entries of the charged lepton Yukawa matrix are populated by the dimension-5 operators $\bar L_i H \phi^\dagger E_i $ for muons and taus, $i=2,3$, and by the dimension-6 operator $\bar L_1 H \phi^2 E_1 $ for the electrons.
This is consistent with the smallness of the charged lepton masses and with the hierarchy between electron and muon and tau Yukawas.
Such higher-dimension operators can be generated, for instance, by integrating out a set of heavy vector-like leptons at tree level with masses $M\gg\langle \phi \rangle$ well above the EW scale. The muon and tau Yukawas are then suppressed by $\langle \phi\rangle/M$ and the electron Yukawa by $(\langle \phi\rangle/M)^2$.

The off-diagonal terms in the Yukawa matrix are predicted to be zero up to corrections from operators of dimension 10 or higher, assuming that the necessary fields required to mediate such off-diagonal operators are even present in the UV theory.
The suppressed mixing provides the needed protection against cLFV constraints.
A phenomenologically viable realization of the neutrino masses and mixings, on the other hand, requires additional scalars.
Thanks to the smallness of the neutrino masses, this can be done consistently without introducing sizeable cLFV.

\begin{figure*}[t]
	\centering
	\includegraphics[width=15cm]{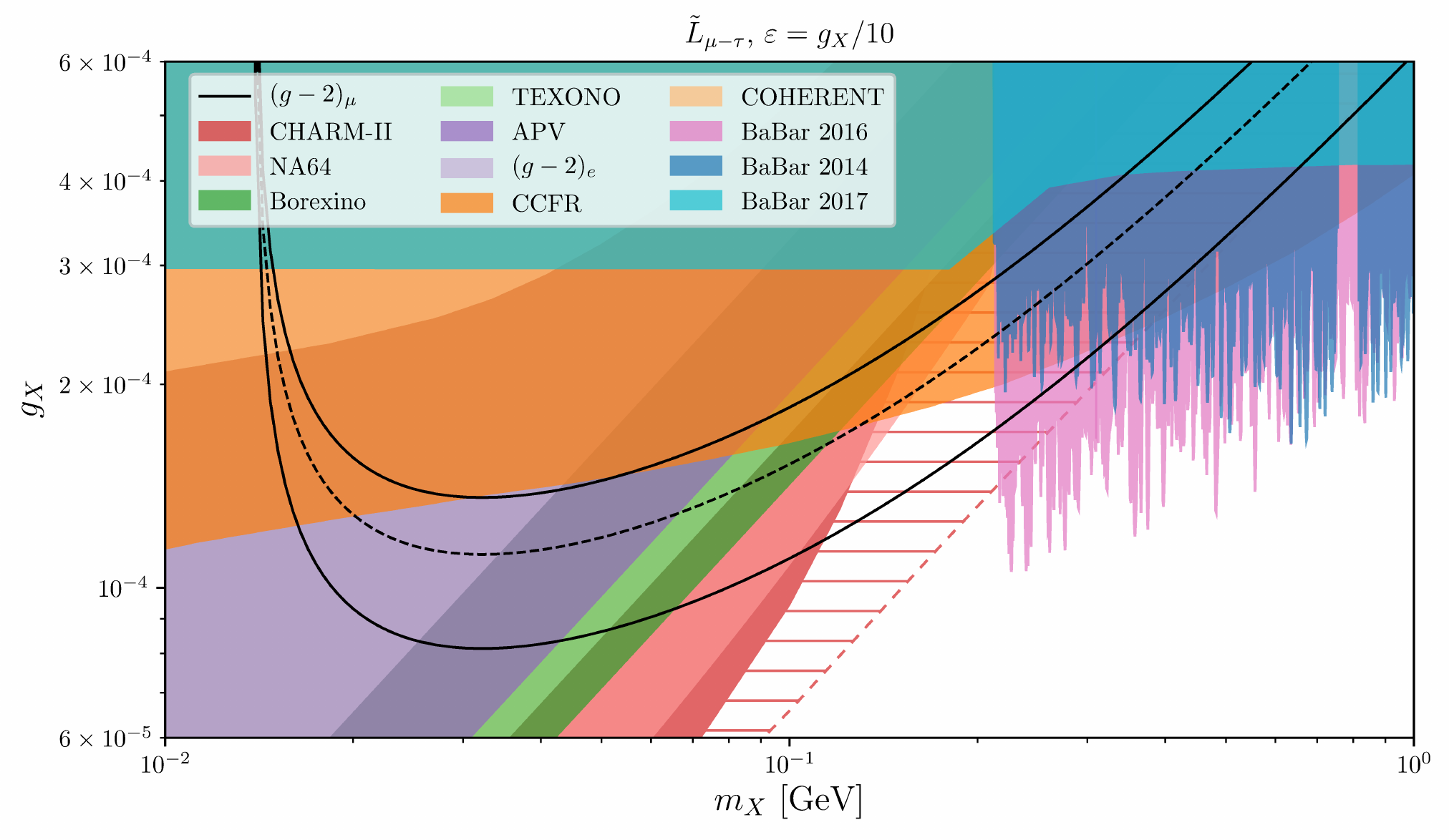}
	\caption{Parameter space for the light $ X $ solution to the $ (g-2)_\mu $ anomaly in the $\tilde{L}_{\mu-\tau}$ model. The shaded regions are excluded by various experiments, while the region between the black lines is preferred by $ (g-2)_\mu $ at $2 \sigma$. See Section~\ref{sec:model-II} for details.} \label{fig:Lt}
\end{figure*}

%%%%%%%%%%%

The $2\sigma$ band in parameter space of $\tilde L_{\mu-\tau}$ that explains \gmu  is denoted with solid-black lines in Fig.~\ref{fig:Lt} (dashed black line denotes the central value), while the colored regions are excluded.
The neutrino trident CCFR bound (orange)
limits the $\tilde L_{\mu-\tau} $ solution of \gmu to $ m_X \lesssim \SI{400}{MeV}$. In Fig.~\ref{fig:Lt} the kinetic mixing parameter was set to $\varepsilon = g_X /10 $, comparable to the IR contribution from muon and tau running in the loop, cf. Eq. \eqref{eq:epsilon:radiative} below. For larger values of $\varepsilon$, the atomic parity violation constraints (dark purple) become more stringent (see Section \ref{sec:APV}).
The upper limit on  $\varepsilon$ from atomic parity violation makes the CCFR bound very robust, i.e., the trident can not be removed by choosing $\varepsilon$ as in the vector category.

The Borexino bound on neutrino--electron scattering (dark green) limits  the $\tilde L_{\mu-\tau} $ solution of \gmu to the $X$ masses above $ m_X \gtrsim \SI{60}{MeV}$. The  NSI oscillation bounds, on the other hand, are not relevant for the  $\tilde L_{\mu-\tau}$ model, since in $\tilde L_{\mu-\tau}$ the couplings of $X_\mu$ to electrons are purely axial and, thus, induce only suppressed, spin-dependent effects. Furthermore,  in $\tilde L_{\mu-\tau}$ the $X$ boson does not couple to quarks. These two features of  $\tilde L_{\mu-\tau}$ couplings mean that the NSI oscillation bounds are completely avoided in this model.

The constraints from the resonant searches are obtained using the \texttt{DarkCast} code which by default supports only vector couplings. We approximate the $\tilde L_{\mu-\tau}$ bounds with those for a vector model with charges $x_{L_1,L_2,L_3} =  (1,\, \frac{13}{2},\, -\frac{13}{2}) $. For processes where the mass can be neglected, such as NA64, there is no difference between the axial and the vectorial couplings. Also for the BaBar search, the above vector model reinterpretation still approximates rather well the actual bounds~\cite{Kahn:2016vjr,axial:DarkCast}. Below the di-muon threshold, the dominate decay channel of $X$ is to invisible final states, while above the di-muon threshold there is a sizeable branching ratio for $X\to \mu^+\mu^-$. The $e^+ e^-$ decay mode is suppressed because of the charge hierarchy and gives a sub-leading constraint. In all cases, the $X$ decays are prompt in the targeted parameter space. The resonance searches  rule out the $\tilde L_{\mu-\tau}$ solution to $(g-2)_\mu$ for $m_X > 2 m_\mu$ (BaBar) and for $m_X \lesssim 100$\,MeV (NA64).

Since $\tilde L_{\mu-\tau}$ has chiral couplings, there are also observables that are only important for such cases of combined vector and axial vector couplings to SM fermions, which we discuss next.

%%%%%%%%%%%
\subsection{Atomic parity violation}
\label{sec:APV}
Parity-violating atomic transitions can determine whether $ X $ couples axially to electrons.
For typical momentum transfers in atomic interactions $ q^2 \ll m_X^2 $ (for the mass range considered here) and the tree-level exchange of $X$ leads to an effective interaction,
    \begin{equation}
    \mathcal{L}_\mathrm{eff} \supset  \dfrac{\varepsilon\, e\, g_X}{m_X^2} (\bar{e} \gamma_\mu \gamma_5 e)  (\tfrac{2}{3} \bar{u} \gamma^\mu u  - \tfrac{1}{3} \bar{d} \gamma^\mu d),
    \end{equation}
where the couplings to quarks are due to kinetic mixing. The effect of  $X$ on the Hamiltonian of atomic systems is, thus, modeled with a parity-violating point-like interaction between the electric charge of the nucleus and electrons and mimics the parity-violating interaction between the nucleus and electrons mediated by the exchange of the SM $ Z $ boson. Hence, the new muonic force can be seen as a modification of the weak charge of the nucleus~\cite{Arcadi:2019uif},
    \begin{equation}
    \Delta Q_\mathrm{weak} = Z\, \varepsilon\, e\, g_X \dfrac{2\sqrt{2} }{G_F m_X^2},
    \end{equation}
where $ Z $ is the atomic number.

The parity-violating transition has been measured to percent-level precision in $ \phantom{}_{55}^{133}\mathrm{Cs} $~\cite{Porsev:2009pr}, with the latest analysis finding $ Q^{\mathrm{exp}}_\mathrm{weak} - Q^{\sscript{SM}}_\mathrm{weak} = \num{.65(43)} $~\cite{Dzuba:2012kx}. We can use this to place the $95\%$ CL bound on the atomic parity violating (APV) couplings of the $ X $ boson,
    \begin{equation}
    \label{eq:APV:bounds}
    g_X <\dfrac{m_X}{\SI{1}{GeV}} \left(\dfrac{|\varepsilon| e}{g_X} \right)^{\!\! \eminus 1/2}
    \begin{dcases}
    \num{3.4e-4} & \mathrm{for}\; \varepsilon >0 \\
    \num{1.3e-4} & \mathrm{for}\; \varepsilon <0
    \end{dcases}.
    \end{equation}

With no particular UV theory in mind, $ \varepsilon $ is treated as a free parameter, and the APV bound can be evaded by sufficiently reducing the value of $\varepsilon$.\footnote{In case of gauge unification, $ \varepsilon $ vanishes above the breaking scale. It is then radiatively generated and calculable.} Nevertheless, the running of $ \varepsilon $ can be used to set an approximate lower bound on $ |\varepsilon| $ in the absence of tuning. Assuming no other BSM particles below the scale of electroweak symmetry-breaking and working in the limit $ \varepsilon \ll g_X$, the 1-loop running of $ \varepsilon $ receives its dominant contribution from renormalization group running between the muon and tau mass~\cite{Greljo:2021npi}, resulting in
    \begin{equation}
    \label{eq:epsilon:radiative}
    \varepsilon(m_\mu) - \varepsilon(m_\tau) \simeq \dfrac{13e\, g_X }{12\pi^2} \ln \dfrac{m_\tau}{m_\mu} = 0.31 e \, g_X.
    \end{equation}
For processes with small momentum transfers, we therefore expect $ \varepsilon \gtrsim g_X/10$.
At high scales, where other states, such as the putative $ S_3 $ muoquark, are dynamical, these can also contribute to the running of $ \varepsilon $, which should be taken into account in any UV completion of the model.

The presence of a rather stringent APV bound, Eq.~\eqref{eq:APV:bounds}, has important phenomenological implications.
It excludes the possibility of the $ X $ coupling to muons being dominated by the kinetic mixing, which could otherwise decorrelate the couplings to muons and muon neutrinos, cf. Section~\ref{sec:NTP}. This is not possible for $\tilde L_{\mu -\tau}$ .
Since in this model the interactions of $X_\mu$ with nucleus are induced by the kinetic mixing, the COHERENT bound in Fig.~\ref{fig:Lt} (light orange) can be directly compared with the one from APV (dark purple), the latter being stronger.

%%%%%%%%%%%
\subsection{Electron anomalous magnetic moment}
The second observable for which the $ \tilde{L}_{\mu -\tau} $ model is qualitatively different from a vector-like model such as $ L_{\mu} -L_{\tau} $ is the electron anomalous magnetic moment, $(g-2)_e$. In $ L_{\mu} -L_{\tau} $ the $ X $ coupling to electrons is exclusively due to a (presumed) small kinetic mixing with the photon, resulting in a similarly minute modification of $ (g-2)_e $. In the $ \tilde{L}_{\mu -\tau} $ model, on the other hand, the electron carries a nonzero $ \U(1)_X $ charge and receives a much larger contributions to $(g-2)_e$ (the contribution is also numerically larger because of the coupling to electrons is axial, cf. Eq. \eqref{eq:amu_NP}).

The $ (g-2)_e$ has been calculated to high precision in the SM~\cite{Aoyama:2014sxa} but crucially depends on the exact value of the fine-structure constant. Recently, $ \alpha $ has been measured precisely  in two experiments involving cesium~\cite{Parker:2018vye} and rubidium~\cite{Morel:2020dww} atoms, respectively; however, the measurements are internally inconsistent at the level of $ 5.4 \sigma $. With these inputs, the deviation of the measured $ (g-2)_e$~\cite{Hanneke:2008tm} from the SM theory prediction is found to be $ a_e^\mathrm{exp}- a_e^{\sscript{SM},\mathrm{Cs}} =
(-8.8\pm 3.6)\cdot 10^{-13} $
 and $ a_e^\mathrm{exp}- a_e^{\sscript{SM},\mathrm{Rb}} =
  (4.8\pm 3.0) \cdot 10^{-13} $,
  with errors dominated by the experimental measurement of $ (g-2)_e$.

In the $\tilde{L}_{\mu-\tau} $ model, the axial coupling of $ X $ to the electron gives a negative correction to $ (g-2)_e $ (see, e.g., Refs.~\cite{Jackiw:1972jz,Jegerlehner:2009ry}). Working in the limit $ m_X \gg m_e $, this translates into 95\% CL bounds
    \begin{equation}
    g_X < \num{9.5e-3} \dfrac{m_X}{\SI{1}{GeV}} (-\Delta a_e) = \begin{dcases}
    \num{1.2e-2} & \text{[Cs]},\\
    \num{3.3e-3} & \text{[Rb]}.
    \end{dcases}
    \end{equation}
A stronger bound is obtained using $ \alpha $ from the rubidium measurement, which prefers a positive contribution to $ (g-2)_e $. However, even this constraint is still weaker than the Borexino bound as shown in Fig.~\ref{fig:Lt}.

\subsection{Neutrino-electron scattering}

The reactor experiments TEXONO~\cite{TEXONO:2009knm} and GEMMA~\cite{Beda:2009kx} measured the neutrino scattering on electrons and set competitive constraints on the $X_\mu$ couplings. The relevant bound shown in Fig.~\ref{fig:Lt} is extracted using the $\bar \nu_e e^- \to \bar \nu_e e^-$ electron recoil energy spectrum from  Fig.~16(b) of the TEXONO analysis paper, Ref~\cite{TEXONO:2009knm}, closely following the procedure described in~\cite{Lindner:2018kjo}. For $\tilde L_{\mu-\tau}$, we find the
$95\%$ CL bounds
    \beq
    g_X < 1.7\times 10^{-4} \frac{m_X}{100~\rm{MeV}}~,
    \eeq
and $2.8\times 10^{-4} < g_X \times 100~{\rm MeV} /m_X< 3.5\times 10^{-4}$, where the NP amplitude is roughly twice the negative SM amplitude. The electron recoil energy is $\lesssim 8$\,MeV, thus, the above EFT limit is valid for the $m_X$ range considered in Fig.~\ref{fig:Lt}. This bound is only slightly worse than the Borexino bound.

The high-energy beam experiment CHARM-II at CERN measured elastic $\nu_\mu e^-$ and $\bar \nu_\mu e^-$ scattering~\cite{CHARM-II:1993phx,CHARM-II:1994dzw}. The target calorimeter was exposed to the horn-focused wide band neutrino beam from the Super Proton Synchrotron with the mean muon neutrino (antineutrino) energy of $E_{\nu} = 23.7 \pm 0.3$\,GeV ($E_{\nu} = 19.1 \pm 0.2$\,GeV).  The signature of this process is a forward-scattered electron producing an electromagnetic shower measured by the calorimeter. The variable $y$ is defined as the ratio of the electron recoil energy over the incident neutrino energy. Neglecting the electron mass, the differential cross section is given by
\beq
\frac{d \sigma}{d y}(\nu_\mu e, \bar \nu_\mu e) = \frac{2  G_F^2 E_\nu m_e}{\pi} \left( g_{\LL,\RR}^2 + g_{\RR,\LL}^2 (1-y)^2 \right)~,
\eeq
where
\begin{align}
g_{\LL} &= s_w^2 - \frac{1}{2} + \frac{g_X^2 x_{L_2} x_{L_1}}{2\sqrt{2} G_F (2 E_\nu m_e y + m_X^2)}  ,\\
g_\RR &= s_w^2 + \frac{g_X^2 x_{L_2} x_{E_1}}{2\sqrt{2} G_F (2 E_\nu m_e y + m_X^2)},
\end{align}
where the $ X$ charges of the $\tilde L_{\mu-\tau}$ model are given in Eq.~\eqref{eq:X_Lt}.

The unfolded differential distributions in $y$ for $\nu_\mu e$ and $\bar \nu_\mu e$ scattering were reported in Fig.~1 of Ref.~\cite{CHARM-II:1993phx}, albeit in arbitrary units. This information was then used in Ref.~\cite{CHARM-II:1993phx} to determine the SM weak mixing angle by fitting to the shape of the distributions, while requiring no knowledge of the overall normalization. Similarly, in $\tilde L_{\mu-\tau}$ we are able to set an upper bound on $g_X$ for given $m_X$ by profiling over two arbitrary nuisance parameters describing the absolute normalisations. The excluded region is shown with red color in Fig.~\ref{fig:Lt}. This represents the most stringent limit on the model for $m_X \lesssim 100$\,MeV.

The shape analysis is unfortunately ineffective for larger masses where the information on the overall rate is needed. Although the SM prediction was overlaid in Fig.~1 of Ref.~\cite{CHARM-II:1993phx}, the uncertainty on the prediction was not reported. The expected SM cross section depends on the number of factors subject to systematic uncertainties such as the neutrino flux determination.  Conservatively assuming the error on the normalisation of the SM prediction to be $30\%$, i.e., constraining the two nuisance parameters  in the fit  to be $1.0\pm 0.3$, we find the exclusion shown with the red-hatched region in Fig.~\ref{fig:Lt}, covering the remaining viable mass window for \gmu. While the estimate of a 30\% error on overall normalization seems plausible to us, if not even conservative (it is an order of magnitude larger than the relative error in the determination of $g_A^{\nu e}$~\cite{CHARM-II:1994dzw}), it is not sufficient to claim a definitive exclusion of the $\tilde L_{\mu-\tau}$ model. A proper recast of the CHARM-II bound using the detector level events shown in Fig.~1 of Ref.~\cite{CHARM-II:1994dzw} while accounting correctly for the systematic effects could possibly achieve this, but it is beyond the scope of this work.

\section{Possible connections to $B$-anomalies}
\label{sec:B-anomalies}
We continue by commenting on the possibility that a single vector mediator is behind a simultaneous solution of $(g-2)_\mu$ and $B$ anomalies. As shown in Section~\ref{sec:generic}, there is a generic upper bound on the vector mass from neutrino trident production and $T$ parameter, $m_X \lesssim 4$\,GeV. In this situation, $B \to K  \nu \bar \nu$ provides an important bound since it receives contributions from on-shell $B \to K X$ decay. Generically, $X$ decays invisibly with a sizeable branching ratio. As shown in Ref.~\cite{Greljo:2021npi}, this puts an upper limit on the effective $g_{b s}$ coupling for a given $m_X$. Since the deviation in $R_{K}$ is proportional to the product of $g_{b s}$ and the effective $g_{\mu \mu}$ coupling, the two together imply a lower bound on $g_{\mu \mu}$ for given $m_X$, restricting the viable parameter space to be in the orange region in Figure~1 of Ref.~\cite{Greljo:2021npi}.  The range predicted for the effective $g_{\mu \mu}$ coupling from the $(g-2)_\mu$ anomaly, on the other hand, does not overlap with the orange band in Figure~1 of Ref.~\cite{Greljo:2021npi}, apart from the small region around $m_X\sim 2.5$\,GeV and $q_A / q_V \simeq -0.4$ (see also, e.g., Ref.~\cite{Crivellin:2022obd}). This possibility is ruled out by the constraints on the trident production in the case of a small kinetic mixing. When the requirement of vanishingly small kinetic mixing is relaxed, this conclusion no longer applies in full generality, since the kinetic mixing can be used to remove the trident constraint. One then needs to invoke constraints from resonance searches to rule out explicit models.

A better way to solve the $B$ anomalies in this framework is to extend the field content by a heavy scalar, a TeV-scale muoquark $S_3$ in the $(\repbar{3},\, \rep{3},\, 1/3)_{x_{S_3}}$ representation of the SM gauge group $ \SU(3)_c\times \SU(2)_\LL \times \U(1)_Y$ and with the $ \U(1)_X $ charge $x_{S_3}=-x_{L_2} - x_q $ for $x_{L_2} \neq x_{L_1, L_3}$. These charge assignments allow  a Yukawa coupling ($\bar{q}^c\ell S_3$) of the muoquark to $\mu$ but not to $e$ and $\tau$. Furthermore,  for $ x_{L_2} \neq - 3 x_q$, the dimension-4 diquark operators $q q S_3^\dagger$ are forbidden and proton decay is suppressed. In other words, a non-universal $\U(1)_X$ enhances the properties of a TeV-scale leptoquark by keeping the accidental symmetries of the SM while still addressing $B$-anomalies~\cite{Greljo:2021xmg,Davighi:2020qqa,Hambye:2017qix,Davighi:2022qgb,Heeck:2022znj}. Out of 276 quark-flavor universal $\U(1)_X$ charge assignments,  273 allow for the above conditions for inclusion of a {muoquark}~\cite{Greljo:2021npi}. The three models which fail are: the dark photon model, which has $x_f = 0$ for all fermions, and the $\U(1)_{B-L}$ and $\U(1)_{N_i-N_j}$ models.
The $b\to s \mu^+\mu^-$ anomaly is resolved by a tree-level exchange of a TeV-scale $S_{3}$ muoquark. This exchange leads to an additional $V-A$ contribution to the $b \to s \mu^+ \mu^-$ transitions, as required by data; see, e.g., Refs.~\cite{Greljo:2021xmg,Davighi:2020qqa,Hiller:2014yaa,Dorsner:2016wpm,Buttazzo:2017ixm,Crivellin:2017zlb,Hiller:2017bzc, Gherardi:2020qhc,Angelescu:2021lln,Marzocca:2018wcf,Dorsner:2017ufx,Babu:2020hun}.

%%%%%%%%%%%%%%%%%%%%%%%%%%%%%%%%%%%
\section{Conclusions}
\label{sec:conc}
%%%%%%%%%%%%%%%%%%%%%%%%%%%%%%%%%%

A new massive spin-1 boson $X_\mu$ coupling vectorially to muons, $\mathcal L \supset g_X q_V \bar \mu \gamma^\mu \mu \, X_\mu$, can give a one-loop contribution of the right size to explain the  $4.2\sigma$ discrepancy in $(g-2)_\mu$. Such a bottom-up simplified model is relatively poorly constrained experimentally. The mass of the $X$ boson can be anywhere from $10$\,MeV (set by cosmological constraints, see Section~\ref{sec:cosmo}) up to $1$\,TeV (the perturbative unitarity limit, see Section \ref{sec:g-2}). Fully exploring this mass window via direct searches may well require a dedicated future collider strategy such as a muon collider~\cite{Capdevilla:2021kcf}.

However, additional theoretical inputs more often than not lead to correlations with  other phenomenological probes. In general, these either severely constrain the $X_\mu$ solution to the $(g-2)_\mu$ anomaly or predict a signal in the next generation of experiments. For instance, demanding electroweak gauge invariance, the renormalizable couplings of $X_\mu$  to muons come from two sources: the kinetic mixing, ${\cal L}\supset \varepsilon_Y B_{\mu \nu} X^{\mu \nu}/2$, and the covariant derivative in the kinetic term, which gives couplings of the form  $\mathcal L \supset g_X \big(x_{L_2} \bar L_2 \gamma^\mu L_2  + x_{E_2} \bar E_2 \gamma^\mu E_2 \big)\, X_\mu$. Each of the two terms comes with a separate set of constraints.
Due to electroweak gauge invariance, the covariant derivative term generates couplings of the $X$ boson not just to muons but also to muon neutrinos. This then leads to constraints from neutrino trident production, which is most relevant in the high $X$ mass region. Meanwhile, the kinetic mixing is constrained by electroweak precision data, which is also mostly relevant for  heavier $ X$ masses starting from around the GeV scale.
Even before committing to a specific gauged $\U(1)_X$ model, the combination of the neutrino trident production and electroweak precision tests, assuming $\U(1)_X$ breaking by SM gauge singlets,  already limits  any such solution of $(g-2)_\mu$ to relatively light $X$ boson masses, $m_X\lesssim 4\,\text{GeV}$, see Section~\ref{sec:generic}.

Further phenomenological implications are obtained in specific complete theories that contain the $X$ boson. In this manuscript, we performed a comprehensive survey of spontaneously broken anomaly-free gauged $\U(1)_X$ models in which the SM matter content is minimally extended by three generations of right-handed neutrinos.
Our results are independent of how the $\U(1)_X$ gauge group is broken, as long as the condensate is neutral under the SM.
In the mass window $10\,\text{MeV}\lesssim m_X\lesssim 4$ GeV we perform a thorough investigation of 419 phenomenologically inequivalent renormalizable models with vector-like charge assignments, allowing for arbitrary kinetic mixing, i.e., for
the full set of quark flavor--universal models that have vector-like $\U(1)_X$ charge assignments for charged SM leptons, and a maximal (finite) $\U(1)_X$ charge ratio of 10.
We find that 7 such models, listed in Table~\ref{tab:viable_models}, avoid the experimental constraints in a narrow mass window with $m_X$ just below $2 m_\mu$ such that the global tension with all the data, including $(g-2)_\mu$, is less than $2\sigma$. The viable models have charge assignments that are either exactly $L_\mu-L_\tau$ or, in most cases, deformations of it in the  $B-2L_e- L_\tau$ direction, which to a large extent avoids the stringent constraints on nonstandard neutrino interactions from neutrino oscillations.  The best agreement with the data is obtained for $L_\mu-L_\tau$. The key complementary constraints on the models are due to the searches for nonstandard neutrino interactions, either the bounds from neutrino oscillations or from COHERENT and Borexino.

There are also 21 chiral anomaly-free $\U(1)_X$ charge assignments with charges in the range $[-10,10]$. The contributions to $(g-2)_\mu$ are maximized for muon couplings that are as close to vector-like as possible, while the effect of bounds on nonstandard neutrino interactions from neutrino oscillations is minimized for axial electron couplings and for vanishing couplings to quarks.  We performed a detailed analysis of a prime candidate of this type, the $\tilde L_{\mu-\tau}$ model, and found that it is excluded in the region relevant for $(g-2)_\mu$. We can reasonably expect that the same is true for the other models with chiral charge assignments.

All of the above $\U(1)_X$ solutions to the $(g-2)_\mu$ anomaly feature lepton flavor--non-universal charge assignments. These imply selection rules on the charged lepton mass matrix and force the interaction and mass bases to coincide up to small, potential corrections from higher-dimensional effective operators. This provides a very effective mechanism to suppress charged lepton flavor violation. Indeed, radiative muon and tau decays in general present the biggest challenge for the new physics explanations of the \gmu anomaly, requiring a rather stringent flavor alignment~\cite{Isidori:2021gqe}. In the above models, this protection against such flavor constraints is built into their symmetry structure. Such setups are also interesting in the context of the ongoing $B$-physics anomalies, allowing for the \textit{muoquark} mediators to be present~\cite{Hambye:2017qix,Davighi:2020qqa,Greljo:2021xmg,Greljo:2021npi,Davighi:2022qgb}.

%%%%%%%%%%%%%%%%%%%%%%%%%%%%
\subsection*{Acknowledgments}

We thank Chaja Baruch and Yotam Soreq for many useful discussions. The work of AG and AET has received funding from the Swiss National Science Foundation (SNF) through the Eccellenza Professorial Fellowship ``Flavor Physics at the High Energy Frontier'' project number 186866.
The work of AG is also partially supported by the European Research Council (ERC) under the European Union’s Horizon 2020 research and innovation programme, grant agreement 833280 (FLAY). The work of PS is supported by the SNF grant 200020\_204075.
JZ acknowledges support in part by the DOE grant de-sc0011784.

\appendix
\section{NSI oscillation bounds} \label{app:NSI}

Here we describe the construction of a $\chi^2$ function that approximates the results of a global fit to the NSI oscillation data~\cite{Coloma:2020gfv}. The starting point is an observation that neutrino oscillations in matter depend on the effective parameters\footnote{%
We only need to consider $\U(1)_X$ couplings that conserve neutrino flavour, and thus the more general coefficients $\mathcal{E}_{\alpha\beta}$ reduce to just the diagonal entries, $\mathcal{E}_{\alpha}=\mathcal{E}_{\alpha\alpha}$.
}  (for more details, see e.g.~\cite{Esteban:2018ppq,Coloma:2020gfv})
\begin{equation}
 \mathcal{E}_{\alpha}(\vec x)
 =
 \varepsilon_{\alpha}^e
 + \varepsilon_{\alpha}^p
 + Y_n(\vec x)\,\varepsilon_{\alpha}^n\,,
\end{equation}
where $\alpha=\{e,\mu,\tau\}$ is the neutrino flavor index, $Y_n(\vec x)$ is the neutron-to-proton ratio in matter at position $\vec x$ along the neutrino trajectory, while $\varepsilon_\alpha^f$ are given by\footnote{We do not include the kinetic mixing contribution to the $ X $ coupling to the fermions, as these contributions cancel in neutral matter.}
\begin{equation}
 \varepsilon_{\alpha}^f = x_f\,x_\alpha\,\varepsilon_0, \qquad f\in\{e,p,n\},\quad \alpha\in\{e,\mu,\tau\},
\end{equation}
where $x_{f,\alpha}$ are the corresponding $\U(1)_X$ charges, and
\begin{equation}\label{eq:NSI_eps0}
 \varepsilon_0 = \frac{1}{\sqrt{2}\,G_F}\frac{g_X^2}{m_X^2}\,.
\end{equation}
The $\U(1)_X$ models we consider have quark flavor--universal couplings, and, thus, the proton and neutron charges are given by the common baryon charge,
\begin{equation}
 x_p = x_n = x_B.
\end{equation}

We make an additional assumption that the approximate $\chi^2$ function can be expressed in terms of the $\vec x$-independent average $\overline{\mathcal{E}}_{\alpha}$, which depends on the averaged $\overline Y_n$ and is given by
\begin{equation}
 \overline{\mathcal{E}}_{\alpha} =
 \left(
  x_e + (1+\overline{Y}_n)\,x_B
 \right) x_\alpha\,\varepsilon_0\,,\qquad \alpha\in\{e,\mu,\tau\}.
\end{equation}
We then define the approximate NSI $\chi^2$ function as
\begin{equation}\label{eq:chi2_NSI}
 \chi^2_{\rm NSI} =
 \left(\vec{\mathcal{E}} - \vec{\mathcal{E}}_0\right)^{\rm T}
 C^{\eminus 1}
 \left(\vec{\mathcal{E}} - \vec{\mathcal{E}}_0\right),
\end{equation}
where
$\vec{\mathcal{E}} = \left(\overline{\mathcal{E}}_{e}, \overline{\mathcal{E}}_{\mu}, \overline{\mathcal{E}}_{\tau}\right)^{\rm T}$,
while
$\vec{\mathcal{E}}_0 = \left(\mathcal{E}_{e}^0, \mathcal{E}_{\mu}^0, \mathcal{E}_{\tau}^0\right)^{\rm T}$
denotes the minimum of the $\chi^2$ function.
The components of the covariance matrix $C$ are given by
\begin{equation}
 C_{\alpha \beta} = \sigma_\alpha\,\sigma_\beta\,\rho_{\alpha\beta}\,,
\end{equation}
with the standard deviations $\sigma_\alpha$ and the correlation coefficients $\rho_{\alpha\beta}$, which satisfy
\begin{equation}
 |\rho_{\alpha\beta}|\leq 1\,,\qquad \rho_{\alpha\beta} = \rho_{\beta\alpha}\,,\qquad \rho_{\alpha\alpha}=1\,.
\end{equation}
The parameters $\vec\eta$ that define the $\chi^2$ function,
\begin{equation}
 \vec\eta =(\mathcal{E}_{e}^0, \mathcal{E}_{\mu}^0, \mathcal{E}_{\tau}^0, \sigma_e, \sigma_\mu, \sigma_\tau, \rho_{e\mu}, \rho_{e\tau}, \rho_{\mu\tau})^{\rm T}\,,
\end{equation}
are then determined from the fit results of Ref.~\cite{Coloma:2020gfv}, which are provided for various model parameters $\vec \theta$,
\begin{equation}
 \vec \theta = (x_B, x_e, x_\mu, x_\tau)^{\rm T}\,.
\end{equation}
The central quantity that was used to set the bounds in~\cite{Coloma:2020gfv} is
\begin{equation}
 \Delta \chi^2_{\rm NSI} = \chi^2_{\rm NSI} - \chi^2_{\rm NSI,\,SM}\,,
\end{equation}
where $\chi^2_{\rm NSI,\,SM}$ is the value of the $\chi^2$ function at the SM point. Using the definition in Eq.~\eqref{eq:chi2_NSI}, along with $\varepsilon_0|_{\rm SM}=0$ and $\vec{\mathcal{E}}|_{\rm SM} = \vec 0$,  we thus have
\begin{equation}\label{eq:Delta_chi2_NSI}
 \Delta \chi^2_{\rm NSI} =
 \left(\vec{\mathcal{E}} - \vec{\mathcal{E}}_0\right)^{\rm T}
 C^{\eminus 1}
 \left(\vec{\mathcal{E}} - \vec{\mathcal{E}}_0\right)
 -
 \vec{\mathcal{E}}_0^{\rm T}\,
 C^{\eminus 1}\,
 \vec{\mathcal{E}}_0\,.
\end{equation}

In order to determine the parameters $\vec\eta$, we use two quantities that were provided in Ref.~\cite{Coloma:2020gfv} for several different models, i.e., for several different values of $\vec \theta$ (see the first column in Table~\ref{tab:chi2_NSI_input}):
\begin{itemize}
 \item The first quantity is the $2\sigma$ bound on $g_X/m_X$ in a given model, $\vec \theta_i$, which is obtained when $\Delta \chi^2_{\rm NSI} =4$. We convert this to a bound on $\varepsilon_0$ using \eqref{eq:NSI_eps0}, giving the $\hat\varepsilon_{0,i}^{\rm bnd}$ values listed in the last column of Table \ref{tab:chi2_NSI_input}. Ideally, these bounds would be reproduced  for each $\vec \theta_i$ by minimizing  the $\Delta \chi^2_{\rm NSI}$ function, Eq. \eqref{eq:Delta_chi2_NSI}, after a judicial choice of the values for nuisance parameters $\vec \eta$ and $\overline{Y}_n$.
 To this end, we solve the equation $\Delta \chi^2_{\rm NSI}(\varepsilon_0, \vec \theta, \vec \eta, \overline{Y}_n) =4$ for $\varepsilon_0$, which results in a function
  \begin{equation}
  \varepsilon_0^{{\rm bnd}}(\vec \theta, \vec \eta, \overline{Y}_n)\,,
 \end{equation}
that depends on the model parameters $\vec \theta$, as well as on the $\chi^2$~parameters $\vec \eta$, and on $\overline{Y}_n$. The nuisance parameters need to be chosen so as to minimize the differences between $\hat\varepsilon_{0,i}^{\rm bnd}$ and  $\varepsilon_0^{{\rm bnd}}$ for each $\vec \theta_i$.

 \item The second set of quantities are  $\Delta \hat\chi^2_{{\rm NSI,min},i}$,  the minimum values of $\Delta \chi^2_{\rm NSI}$ for each model $\vec \theta_i$, listed in the second column of Table \ref{tab:chi2_NSI_input}. If the nuisance parameters are chosen correctly, $\Delta \hat\chi^2_{{\rm NSI,min},i}$ should be well reproduced by
  \begin{equation}
  \Delta \chi^2_{\rm NSI,min}(\vec \theta_i, \vec \eta, \overline{Y}_n)\,.
 \end{equation}
Here $\Delta \chi^2_{\rm NSI,min}(\vec \theta, \vec \eta, \overline{Y}_n)$ is a function obtained by minimizing Eq.~\eqref{eq:Delta_chi2_NSI} with respect to $\varepsilon_0$.

 \begin{table}[t]
\setlength{\tabcolsep}{0.2em}
\vspace{0.2cm}
\begin{center}
\begin{tabular}{rrrrcc}
\hline\hline
\multicolumn{4}{c}{${\vec\theta_i}^{\,T}$} & ~~$\Delta \hat\chi^2_{{\rm NSI,min},i}$~~ & ~~~$\hat\varepsilon_{0, i}^{\rm bnd}$~~~ \\
\hline
$(1,$&$-3,$&$0,$&$0)$~ & $-1.4$ & $0.242$ \\
$(1,$&$0,$&$-3,$&$0)$~ & $0$ & $0.0128$ \\
$(1,$&$0,$&$0,$&$-3)$~ & $-0.6$ & $0.0134$ \\
$(1,$&$0,$&$-2/3,$&$-2/3)$~ & $-1.1$ & $0.293$ \\
$(0,$&$1,$&$-1,$&$0)$~ & $-1.3$ & $0.0873$ \\
$(0,$&$1,$&$0,$&$-1)$~ & $-1.0$ & $0.0873$ \\
$(0,$&$ 1,$&$ -1/2,$&$~ -1/2)$ & $-1.3$ & $0.546$ \\
$(1,$&$ 0,$&$ 1,$&$ 1)$~ & $0$ & $0.136$ \\
$(0,$&$ 1,$&$ 2,$&$ 2)$~ & $-0.1$ & $0.196$ \\
$(1,$&$ -2,$&$ -1,$&$ 0)$~ & N/A & $1.58$ \\
\hline\hline
\end{tabular}
\end{center}
\caption{Input data from Ref.~\cite{Coloma:2020gfv} used for fitting the parameters of the approximate $\Delta \chi^2_{\rm NSI}$.
The value of $\hat\varepsilon_{0, i}^{\rm bnd}$ for ${\vec\theta_i}^{\,T} = (1, -2, -1, 0)$ has been provided by the authors of~\cite{Coloma:2020gfv} in private communication in the context of~\cite{Greljo:2021npi}.
}
\label{tab:chi2_NSI_input}
\end{table}

\end{itemize}
To find the best values for nuisance parameters we construct a loss function $\lambda$,
\begin{equation}
 \lambda(\vec \eta, \overline{Y}_n) = \sum_i \left(
\frac{
\left(\varepsilon_{0}^{\rm bnd}(\vec \theta_i, \vec \eta, \overline{Y}_n) - \hat\varepsilon_{0, i}^{\rm bnd}
\right)^2
}{\sigma_{\varepsilon,i}^2}
+
\frac{
\left(\Delta \chi^2_{\rm NSI,min}(\vec \theta_i, \vec \eta, \overline{Y}_n) - \Delta \hat\chi^2_{{\rm NSI,min},i}
\right)^2
}{\sigma_{\chi,i}^2}
 \right)\,,
\end{equation}
where $i$ is an index labelling the different models with parameters $\vec \theta_i$ for which~\cite{Coloma:2020gfv} provides the bounds $\hat\varepsilon_{0,i}^{\rm bnd}$ and the minimum values $\Delta \hat\chi^2_{{\rm NSI,min},i}$, listed in Table~\ref{tab:chi2_NSI_input}.
The standard deviations $\sigma_{\varepsilon\,i}$ and $\sigma_{\chi\,i}$ are chosen as follows:
\begin{itemize}
 \item In reproducing the $\hat\varepsilon_{0,i}^{\rm bnd} $ bounds  we allow a nominal 10\% uncertainty on the value of the approximate $\chi^2$ function, motivated by how large we expect the deviations between the exact and the approximate $\chi^2$ values to be.
 We thus set
 \begin{equation}
  \sigma_{\varepsilon, i}=0.1\, \hat\varepsilon_{0,i}^{\rm bound}\,.
 \end{equation}
 \item The exact $\chi^2$ function used in Ref.~\cite{Coloma:2020gfv} to obtain the minima $\Delta \hat\chi^2_{{\rm NSI,min},i}$ is not Gaussian. We therefore do not expect the approximate $\chi^2$ function to be able to reproduce the values of minima with high accuracy.
 In fact, we observe that choosing too small values of $\sigma_{\chi, i}$ results in a rather poor fitted $\hat\varepsilon_{0,i}^{\rm bnd}$ values.  Allowing for very large $\sigma_{\chi,i}$ uncertainties, on the other hand, results in a fit not converging at all. As a compromise, we use the constant values
 \begin{equation}
  \sigma_{\chi,i} = 2\,.
 \end{equation}
\end{itemize}
\begin{table}[t]
\setlength{\tabcolsep}{0.6em}
\vspace{0.2cm}
\begin{center}
\begin{tabular}{cccccccccc}
\hline\hline
$\mathcal{E}_{e}^0$ & $\mathcal{E}_{\mu}^0$ & $\mathcal{E}_{\tau}^0$ & $\sigma_e$ & $\sigma_\mu$ & $\sigma_\tau$ & $\rho_{e\mu}$ & $\rho_{e\tau}$ & $\rho_{\mu\tau}$ & $\overline{Y}_n$ \\
\hline
$-1.62$ & $-10.00$ & $-11.15$ & $+5.56$ & $+8.51$ & $+9.59$ & $+0.78$ & $+0.79$ & $+1.00$ & $+1.04$ \\
\hline\hline
\end{tabular}
\end{center}
\caption{Fitted values of the parameters of the approximate $\Delta \chi^2_{\rm NSI}$.}
\label{tab:chi2_NSI_fit}
\end{table}
We minimize the loss function $\lambda$ using the \texttt{iminuit}~\cite{iminuit,James:1975dr} Python package and obtain a good fit with $\lambda|_{\rm min} = 1.43$. This indicates that the approximate $\chi^2$ function reproduces the bounds $\hat\varepsilon^0_{{\rm bound}, i}$ with a better accuracy than the assumed tentative uncertainty of 10\%.
The results of this fit is shown in Table~\ref{tab:chi2_NSI_fit}. Note that the fitted average neutron-to-proton ratio $\overline{Y}_n$ in Table~\ref{tab:chi2_NSI_fit}  is very close to the neutron-to-proton ratio averaged over all the Earth $\overline{Y}_n^{\oplus}=1.051$ (but one does not expect to match it exactly, since  the oscillation data also include results from oscillation inside the sun).

\section{Global $\chi^2$ function}
\label{app:global:chi2}

In Section~\ref{sec:NSI} we use a $\chi^2$ function that combines bounds from various measurements:
    \begin{equation}
    \Delta \chi^2 = \Delta \chi^2_{a_\mu}+\Delta \chi^2_{\rm Borex.}+\Delta \chi^2_{\rm NSI~osc.}+\Delta \chi^2_{\rm COHERENT}+\Delta \chi^2_{\rm reson.}.
    \end{equation}
The contribution $\Delta \chi^2_{a_\mu}$ is defined with respect to the ideal NP model, which would be able to exactly reproduce the central value of $\Delta a_\mu$,
\beq
\Delta \chi^2_{a_\mu}= \left( \frac{\Delta a_\mu\big|_{\rm NP}-\Delta a_\mu\big|_{\rm exp}}{\sigma (\Delta a_\mu)} \right)^{\!\! 2}.
\eeq
Here $\Delta a_\mu\big|_{\rm NP}$ is the shift in the anomalous magnetic moment of the muon due to NP, while $\Delta a_\mu\big|_{\rm exp}=251\times 10^{-11}$ is the difference between the measured value and the consensus SM predictions, with $\sigma(\Delta a_\mu)=59\times 10^{-11}$ the corresponding error (see also Section~\cite{Muong-2:2021ojo}). For the SM $\Delta a_\mu\big|_{\rm NP}=0$, and thus $\Delta \chi^2_{a_\mu}=17.8$.

The contribution $\Delta \chi^2_{\rm Borex.}$ is due to possible NP contributions to the cross section for $^7$Be solar neutrinos scattering on electrons. Averaging the high and low metallicity standard solar model predictions for the solar neutrino fluxes, treating the difference to the two predictions as the systematic error, and adding it in quadrature to the experimental error on the measured Borexino rate, gives $R_{\text{Be}}=\sigma_{\rm Borexino}/\sigma_{\rm SM}=1.06\pm0.09$ for the ratio of the measured and SM neutrino cross sections. From this we construct
\beq
\Delta \chi^2_{\rm Borex.}= \left(\frac{R_{\text{Be}}\big|_{\rm NP}-R_{\text{Be}}\big|_{\rm exp}}{\sigma (R_{\text{Be}})} \right)^{\!\! 2},
\eeq
where $R_{\text{Be}}\big|_{\rm NP}=\sigma_{\rm SM+NP}/\sigma_{\rm SM}$, while $R_{\text{Be}}\big|_{\rm exp}$ and $\sigma (R_{\text{Be}})$ are the central value and the error of the  $R_{\text{Be}}$ measurement, respectively. For the SM $\Delta \chi^2_{\rm Borex.}=0.44$.

For NSI oscillations, we use  $ \Delta \chi^2_{\rm NSI}$ described in Appendix~\ref{app:NSI}. Since this is only an approximation to the true $\chi^2$ for NSI oscillation bounds,  $ \Delta \chi^2_{\rm NSI}$ can become negative in regions where the approximations in Appendix~\ref{app:NSI} become less reliable. We thus take $ \Delta \chi^2_{\rm NSI~osc.} = \Delta \chi^2_{\rm NSI}$  if $ \Delta \chi^2_{\rm NSI}>0$, but set $ \Delta \chi^2_{\rm NSI~osc.} =0$ if $\Delta \chi^2_{\rm NSI}$ becomes negative (signaling the breakdown of our approximations). For the SM $ \Delta \chi^2_{\rm NSI~osc.}=0$.

The term $\Delta \chi^2_{\rm COHERENT}$ encodes the constraints from the COHERENT measurement. To obtain its value we use the code accompanying Ref.~\cite{Denton:2020hop}, and calculate the $\chi^2$ statistics for the 12T,12E binning. We define $\Delta \chi^2_{\rm COHERENT}$ as the difference between the $\chi^2$ value for a particular model and the SM $\chi^2$ value. For SM therefore $\Delta \chi^2_{\rm COHERENT}=0$.

For all the models we require that they pass the bounds on resonance searches as implemented in \texttt{DarkCast}. The term $\Delta \chi^2_{\rm reson.}$ is therefore set to zero, if the model passes the  NA64 and BaBar 2014 bounds, and it set to a very high value if they do not (in the code we use the numerical value $\Delta \chi^2_{\rm reson.}=100$ in that case). For the SM $\Delta \chi^2_{\rm reson.}=0$.

\bibliography{g2biblio}

\end{document}

%% file: PreRamble.tex
\usepackage[english]{babel}
\usepackage{microtype,xspace}
\usepackage[utf8]{inputenc}	% Allows for writing special charachters in the tex-file 

\usepackage{amsfonts,amsmath,amssymb,bm,mathrsfs,mathtools,dsfont} 	% Standard mathematics 
\allowdisplaybreaks 	%Allows pagebreak during multiline math enviroments
\usepackage{xcolor}
\usepackage{hyperref}
\usepackage{slashed}

\usepackage{siunitx}
\usepackage{graphicx}
\graphicspath{{./Figures/}}
\usepackage{multirow}

\usepackage{tabularx}
\newcolumntype{Y}{>{\centering\arraybackslash}X}

\usepackage[shortlabels]{enumitem}
\setlist{itemsep=.1em,topsep=.5em}
\SetEnumerateShortLabel{i}{\textit{\roman*}}

\definecolor{red}{rgb}{0.6,.0706,.1373}
\definecolor{blue}{rgb}{0,0.396,0.741}
\colorlet{blueRef}{blue!80!black}
\colorlet{blueLink}{blue!80!black}
\hypersetup{
	colorlinks, 
	bookmarksopen, 
	bookmarksnumbered,
	citecolor=blueRef, 		%color of links to bibliography
	linkcolor=blueLink,	%color of internal links
	urlcolor=blueLink,			%color of external links 
}

\DeclareMathVersion{sans}
\SetSymbolFont{operators}{sans}{OT1}{cmbr}{m}{n}
\SetSymbolFont{letters}{sans}{OML}{cmbrm}{m}{it}
\SetSymbolFont{symbols}{sans}{OMS}{cmbrs}{m}{n}
\SetMathAlphabet{\mathit}{sans}{OT1}{cmbr}{m}{sl}
\SetMathAlphabet{\mathbf}{sans}{OT1}{cmbr}{bx}{n}
\SetMathAlphabet{\mathtt}{sans}{OT1}{cmtl}{m}{n}
\SetSymbolFont{largesymbols}{sans}{OMX}{iwona}{m}{n}

\DeclareMathVersion{boldsans}
\SetSymbolFont{operators}{boldsans}{OT1}{cmbr}{b}{n}
\SetSymbolFont{letters}{boldsans}{OML}{cmbrm}{b}{it}
\SetSymbolFont{symbols}{boldsans}{OMS}{cmbrs}{b}{n}
\SetMathAlphabet{\mathit}{boldsans}{OT1}{cmbr}{b}{sl}
\SetMathAlphabet{\mathbf}{boldsans}{OT1}{cmbr}{bx}{n}
\SetMathAlphabet{\mathtt}{boldsans}{OT1}{cmtl}{b}{n}
\SetSymbolFont{largesymbols}{boldsans}{OMX}{iwona}{bx}{n}

\usepackage[noindentafter]{titlesec} %Removes the indentation after new section
\titleformat{\section}{\large \bfseries \sffamily \mathversion{boldsans} \color{blue!70!black} }{\thesection}{10pt}{}{}
\titlespacing{\section}{0pt}{10pt}{5pt}
\titleformat{\subsection}{\sffamily \mathversion{sans} \color{blue!80!black} }{\thesubsection}{10pt}{}{}
\titlespacing{\subsection}{0pt}{8pt}{3pt}
\titleformat{\subsubsection}{\normalsize \itshape \sffamily \mathversion{sans} \color{blue!80!black} }{\thesubsubsection}{10pt}{}{}
\titlespacing{\subsubsection}{0pt}{8pt}{2pt}

\renewcommand{\thesection}{\arabic{section}}
\renewcommand{\thesubsection}{\thesection.\arabic{subsection}}
\renewcommand{\thesubsubsection}{\thesubsection.\arabic{subsubsection}}
\makeatletter
\renewcommand{\p@subsection}{}
\renewcommand{\p@subsubsection}{}
\makeatother

\makeatletter
\let\MyIntOrig\int
\def\MyIntSpace{\hspace{-.35em}} %% Configure as needed.
\def\int{\MyInt}
\def\MyInt{\MyIntOrig\MyIntSkipMaybe}
\def\MyIntSkipMaybe{
	\@ifnextchar_{\MyIntSkipScript}{%
		\@ifnextchar^{\MyIntSkipScript}{%
			\@ifnextchar\limits{\MyIntSkipTok}{%
				\@ifnextchar\nolimits{\MyIntSkipTok}{%
					\MyIntSpace}}}}%
}
\def\MyIntSkipScript#1#2{#1{#2}\MyIntSkipMaybe}
\def\MyIntSkipTok#1{#1\MyIntSkipMaybe}

\newcommand{\pushright}[1]{\ifmeasuring@#1\else\omit\hfill$\displaystyle#1$\fi\ignorespaces}
\makeatother